\documentclass[11pt]{article}

 \usepackage{amscd,amsmath,amsfonts,amssymb,amsthm,cite,mathrsfs,color} 
\usepackage{dsfont} \usepackage{leftidx}
\usepackage{tikz} \usepackage{enumerate} 
 \usepackage{appendix}

 \usepackage[left=2.5cm, right=2.5cm, top=2.5cm, bottom=2.5cm]{geometry}
\usepackage{graphicx} \usepackage{xcolor}
\usepackage[colorlinks,linkcolor=red]{hyperref}
\numberwithin{equation}{section} 

\usepackage{amsthm}

\theoremstyle{plain}
\newtheorem{theorem}{Theorem}[section]
\newtheorem{lemma}[theorem]{Lemma}

\newtheorem{proposition}[theorem]{Proposition}

\theoremstyle{definition}
\newtheorem{definition}[theorem]{Definition}

\newtheorem{remark}[theorem]{Remark}

\begin{document}
\title{\bf{ 
Mean eigenvector self-overlap in deformed complex Ginibre ensemble
}}

\author{Lu Zhang\footnotemark[1]}
\renewcommand{\thefootnote}{\fnsymbol{footnote}}
\footnotetext[1]{School of Mathematical Sciences, University of Science and Technology of China, Hefei 230026, P.R.~China. E-mail: zl123456@mail.ustc.edu.cn}


 \maketitle
 
 \begin{abstract}
Consider a  random matrix  of size $N$ as  an additive deformation of  the complex  Ginibre ensemble under a deterministic matrix $X_0$ with a finite rank, independent of $N$. 
We prove that microscopic statistics for the mean diagonal overlap, near the edge point, are characterized by the iterative  erfc integrals, which is first introduced in \cite{LZ12} and only depend on the geometric multiplicity of certain eigenvalue of $X_0$.
We also investigate the microscopic statistics for the mean diagonal overlap of the outlier eigenvalues. Further we get a phenomenon of the phase transition for the mean diagonal overlap, with respect to the modulus of the eigenvalues of $X_0$.

 \end{abstract}

\section{Introduction and main results}

\subsection{Introduction}
Throughout this paper we will consider the deformed complex Ginibre ensembles, which are defined as follows.
 \begin{definition}  \label{GinU} 
A random  complex  
$N\times N$ matrix $X$, is said to belong to the deformed complex Ginibre ensemble with mean matrix 
\begin{equation}\label{meanmatrix}
X_0={\rm diag}\left( A_0,0_{N-r} \right)
\end{equation}
and  time parameter $\tau>0$, denoted by GinUE$_{N}(A_0)$,  if  the joint probability density function for  matrix entries  is given by   
\begin{equation}\label{model}
P_{N}(A_0;X)=    \Big(\frac{N}{\pi\tau}\Big)^{N^2}\
e^{ -\frac{N}{\tau} {\rm Tr} (X-X_0)(X-X_0)^*}.
\end{equation}
 Here $A_0$ is a $r\times r$ constant  matrix. 
 \end{definition}
  
In Definition \ref{GinU}, with probability 1, we can assume that all eigenvalues of the random matrix $X$ have multiplicity 1. Then $X$ is diagonalizable
by a similarity transformation: $X=S Z S^{-1},$ where $Z={\rm diag}\big(
z_1,\cdots,z_N
\big).$ For $i=1,\cdots,N$, denote the right eigenvector $\vec{r}_i$, which is the $i$-th column of $S$ and satisfies $X\vec{r}_i=z_i \vec{r}_i$; Also denote the left eigenvector $\vec{l}_i^*$, which is the $i$-th row of $S^{-1}$ and satisfies $\vec{l}_i^*X=z_i \vec{l}_i^*$. Note that we can always choose $\vec{r}_i$ and $\vec{l}_i^*$ to satisfy the bi-orthonormality condition: $\vec{l}_i^*\vec{r}_j=\delta_{i,j}$ for $i,j=1,\cdots,N.$ While in general, $S$ is not unitary, so $\vec{r}_i^*\vec{r}_j\not=\delta_{i,j}$ and $\vec{l}_i^*\vec{l}_j\not=\delta_{i,j}$. Then the simplest informative object characterizing the eigenvector non-orthogonality is the so-called `overlap matrix' $\mathcal{O}_{i,j}=(\vec{l}_i^*\vec{l}_j)(\vec{r}_j^*\vec{r}_i)$. In particular, the diagonal entries $\mathcal{O}_{i,i}$ are highly related with the eigenvalue condition numbers in numerical analysis, see e.g. \cite[Sections 35\& 52]{TE}, or \cite{Fy18,BD20,COW2024} for a quick review.

In this paper we will consider the following mean eigenvector self-overlap:
\begin{equation}\label{OverlapStatistic}
\mathcal{O}_N\big(
\tau,A_0,z
\big)=
\frac{1}{N}
\sum_{i=1}^N\mathbb{E}\Big(
 \mathcal{O}_{i,i}\delta(z-z_i)
\Big),
\end{equation}
where the expectation is over the density \eqref{model}. Relative study started with the seminal work of Chalker and Mehlig\cite{CM1998,CM2000,MC1998}, where they estimated the large $N$ limit of the expectation of diagonal and off-diagonal overlaps, for the non-perturbed complex Ginibre ensemble.
 
 The study of Ginibre ensembles for $A_0=0$ in \eqref{model} was first initiated  by  Ginibre  \cite{Gi}. For the eigenvalue statistics, especially for the perturbed case, in the last decade, considerable essential results have been obtained, see e.g. \cite{Ta13,BC16,BZ,BR,COW,OR,MO,LZ12,LZ23,LZ24}, see also \cite{BF23a,BF23b} for a review. 

Also for the eigenvector statistics, in the last decade, considerable progress has been made in the mathematical community, see e.g. 
\cite{WS2015,Mova2016,Fy18,BZ2018,BD20,ATTZ2020,JSS2021,BKMS2021,EKY2023,EJ2023,
CEHS2024,DYYY2024,Osman2024,ABN2024}. A first impressive and important result is the distribution of the diagonal overlap for the complex Ginibre ensemble, see \cite[Theorem 1.1]{BD20} and \cite[Theorem 2.3]{Fy18}; \cite[Theorem 2.1]{Fy18} also gave the distribution of the diagonal overlap for the real eigenvalue of the real Ginibre ensemble. \cite{WS2015} provided a connection between overlaps and mixed matrix moments. It was shown that one and two point functions of eigenvector overlaps, conditioned on an arbitrary number of eigenvalues in the GinUE, lead to determinantal structures\cite{ATTZ2020}, and very recently a Pfaffian structure of the eigenvector overlap for the GinSE has been observed\cite{ABN2024}. For the deformed real or complex i.i.d. matrices which are huge generalisation of \eqref{model}, both lower and upper bounds have been provided for the diagonal overlap \eqref{OverlapStatistic}, see \cite{CEHS2024,EJ2023}. Very recently, for the left and right eigenvectors of the complex i.i.d. matrices, universal results have been obtained \cite{DYYY2024,Osman2024}.

In the physical community, statistical properties of overlaps for non-normal matrices have attracted growing interest in theoretical physics, see e.g. \cite{Fy18,BD20,WCF24} and references therein. One strong motivation comes
from the field of quantum-chaotic scattering, where the statistics of resonances and
phase-shifts could be adequately described by an effective non-Hermitian random matrix approximation $\mathcal{H}_{\rm eff}$, on a scale comparable with the typical level spacing, and essential results have been obtained both for eigenvalue and eigenvector statistics, see e.g. \cite{FM2002,SFPB00,FS1997,FS2003,RI2009}. 

On the other hand, as suggested by Chalker and Mehlig\cite{CM2000}, from the experience of universality in random Hermitian problems, studying of Ginibre ensembles may provide a widely-applicable guide to behaviour, for non-Hermitian operators in a finite system and in an appropriate basis. See also \cite{MS2001} for a hint of the universality for the eigenvector correlations, between Ginibre ensemble and other non-Hermitian ensembles.

In a series of research \cite{JNNPZ1999,BGNTW2014,BGNTW2015,BNST2017,GW2018,NT2018}, with the help of the generalized Green's function for non-hermitian random matrix models in the limit $N\rightarrow +\infty$, one can compute the asymptotic behaviours of the diagonal overlap statistics. This method is based on the two dimensional electrostatic analogy, see e.g. \cite{FS1997,JNNPZ1999}. For a mathematical setting, we refer the reader to \cite[Appendix A]{BD20}. In particular, this method applies well to the complex Ginibre ensemble. Initially, \cite{JNNPZ1999} considered exactly the model \eqref{model} with the spiric case, where the mean matrix $X_0$ is a two level deterministic Hamiltonian, with $N/2$ levels $a$ and $N/2$ levels $-a$, and gave a prediction for the diagonal overlap statistics; By further developing the Burger dynamics, for the model 
\eqref{model}, \cite{BGNTW2015} found an exact formula to compute the macroscopic behaviour of the diagonal overlap statistics, inside the bulk, see  \cite[eqn (39)]{BGNTW2015}; For normal perturbation $X_0$, \cite{GW2018} extended the above result by finding an exact formula of the diagonal overlap statistics, see \cite[eqn (51)]{GW2018}; By using this formula, the authors gave the the macroscopic behaviour in the bulk regime, and in the spiric case, at the collision time, they got an microscopic result which they called the ``collision microscopic law'', see \cite[eqn (61)]{GW2018}. Under this methodology, the overlap statistics have also been investigated for the bi-unitarily invariant ensembles, see e.g. \cite{BNST2017,NT2018}.

There are another important series of research, which focus on the non-perturbed Ginibre and elliptic ensembles, see e.g. \cite{Fy18,FT2021,WCF24,COW2024,CW2024}, with different methodology, which mainly rely on the partial Schur decomposition, and the computation of expectation for product and ratio of determinants. Methodology of our present paper can also be included in this frame work. In this paper, we extend the scope to the case of perturbed complex Ginibre ensembles, where in \eqref{model}, we choose $X_0$ to be an arbitrary non-normal matrix independent of $N$. 

In Proposition \ref{IntegralRepresent}, we give an explicit integration formula for the mean self-overlap defined in \eqref{OverlapStatistic}, for the model \eqref{model} with arbitrary $X_0$ such that $r\leq N-1$; Based on this formula, we observe that under finite rank perturbation, the microscopic statistic of the mean self-overlap near the edge point $z_0$($|z_0|=\sqrt{\tau}$) has an order $O(\sqrt{N})$, and is described by the iterative  erfc functions(cf. \eqref{IEF}) which are first introduced in \cite{LZ12}. In Theorem \ref{OverlapStaEdge}, we observe that the asymptotic statistic only depends on a non-negative integer $t$, which corresponds to the geometric multiplicity if the observing point $z_0$ is an eigenvalue of $X_0$, otherwise it is set to  be zero.

Also in this paper, we investigate the microscopic statistics for the mean self-overlap near the outlier points, in which the asymptotic positions of the outliers have been characterized exactly, and the relative microscopic eigenvalue statistics have been well understood, see \cite{Ta13,BC16,BR}. Different from the edge case, in Theorem \ref{OverlapStaEdgeOutlier1} and Theorem \ref{OverlapStaEdgeOutlier2}, we observe that the microscopic statistic of the mean self-overlap near the outlier points has an order $O(1)$, and is described by a new kind of function, which is different from the edge case. 

The integration formula in Proposition \ref{IntegralRepresent} extends the result in \cite[eqn (51)]{GW2018}, which was limited to the normal perturbation $X_0$; Also, by using 
 this integration formula, we expect to investigate the microscopic statistics at the edge for non-normal infinite rank perturbations, which may thus extend the results in \cite{BGNTW2015}.

\subsection{Main results}
In this paper, for a complex variable $z:=\Re z+{\rm i}\Im z$, we denote its volume element ${\rm d}z$ on the complex plane as ${\rm d}\Re z{\rm d}\Im z.$

To state our main result, we need to introduce  the  Jordan canonical  form of $A_0$. Given  a complex number $\theta$,  
a Jordan block  $R_{p}( \theta)$    is  a $p\times p$  upper triangular matrix of the form
\begin{small}
\begin{equation}\label{1Rpirealedge}
R_{p}\left( \theta\right)= 
\begin{bmatrix}
\theta & 1 & \cdots & 0 \\  & \ddots & \ddots & \vdots \\
&& \ddots & 1 \\ &&& \theta
\end{bmatrix}. 
\end{equation}
\end{small}
Let $z_0$ be the spectral parameter, that is, we will investigate the microscopic behaviour of \begin{small}
$\mathcal{O}_N\big(\tau,A_0,z\big)$
\end{small} near the point $z_0$.
If $z_0$ is an eigenvalue of $A_0$, introduce a non-negative integer $m$, and several non-negative integer pairs $(n_i,p_i)$ such that
\begin{equation} \label{porder}
 p_{1}<p_{2}<\cdots <p_{m}, 
 \end{equation}
which correspond to the distinct sizes of the blocks associated with $z_0$, such that  $R_{p_{i}}( z_0)$ appears  $n_i$ times for $i=1, \ldots, m$. Denote $t$ the geometric multiplicity of $A_0$ with respect to $z_0$ by
\begin{equation}\label{geometricmultiplicity}
t=\sum_{i=1}^m n_i,\quad
r_0=r-\sum_{i=1}^m n_i p_i.
\end{equation}
Otherwise, if $z_0$ is not an eigenvalue of $A_0$, we can simply set $t=0.$ 

Hence, for a certain complex   invertible matrix $P$,     the Jordan canonical  form as
  a direct sum of block diagonal matrices exists
\begin{equation}\label{Jdetail}
A_0=PJP^{-1},\quad
J=\bigoplus_{i=1}^m 
\overbrace{ R_{p_{i}}\left(z_0 \right)   \bigoplus \cdots \bigoplus  
R_{p_{i}}\left(z_0 \right)
}^{n_{i}\ \mathrm{blocks}}\bigoplus J_1,
 \end{equation}
here $J_1$ is of size $r_0\times r_0$ and does not have $z_0$ as an eigenvalue.

 Also, denote  $I_{1}$ ($I_{2}$) the  index set consisting of  numbers corresponding to   first (last)  columns of all 
 blocks  
 $R_{p_{i}}\left(z_0 \right)$(cf. \eqref{1Rpirealedge})    from the Jordan form  $J$ where  $ 1\leq i \leq m$.  Both $I_{1}$ and $I_{2}$ have  the same  cardinality $t=\sum_{i=1}^m n_i$.

Firstly, we provide an integration formula for the mean self-overlap defined in \eqref{OverlapStatistic}.
 \begin{proposition}\label{IntegralRepresent}
For $z=z_0+N^{-\rho}\hat{z}$ and $r\leq N-1$ in \eqref{meanmatrix}, we have
\begin{equation}\label{IntegralRepresentEqua}
\mathcal{O}_N\big(
\tau,A_0,z
\big)=D_N\int_{\overrightarrow{\eta}U\overrightarrow{\eta}\leq 1}
e^{Nf(\overrightarrow{\eta},y)}g(\overrightarrow{\eta},y){\rm d}\overrightarrow{\eta}{\rm d}y,
\end{equation}
where
\begin{equation}\label{U}
U=P^*P,
\end{equation}
\begin{small}
\begin{multline}\label{f eta y}
f(\overrightarrow{\eta},y)=-\frac{1}{\tau}\overrightarrow{\eta}^*
\big( z_0\mathbb{I}_r-J \big)^*U\big( z_0\mathbb{I}_r-J \big)\overrightarrow{\eta}
+\log\big( 1-\overrightarrow{\eta}^*U\overrightarrow{\eta} \big)
\\
+\frac{|z_0|^2}{\tau}\overrightarrow{\eta}^*U\overrightarrow{\eta}+\frac{1}{\tau}
N^{-\rho}\big( \hat{z}\overrightarrow{\eta}^*J^*U\overrightarrow{\eta}
+\overline{\hat{z}}\overrightarrow{\eta}^*UJ\overrightarrow{\eta}
 \big)-\frac{1}{\tau}|y|^2+\log\big(|z|^2+|y|^2\big),
\end{multline}
\begin{equation}\label{DN}
D_N=\frac{\pi^{N-r-1}}{N(N-r-1)!}\big(
N/(\pi \tau)
\big)^{N+1}e^{-\frac{N}{\tau}|z|^2}\det(U)
\end{equation}
\end{small}
and
\begin{small}
\begin{equation}\label{g eta y}
g(\overrightarrow{\eta},y)=
\big(
1-\overrightarrow{\eta}^*U\overrightarrow{\eta}
\big)^{-r-1}
\big(|z|^2+|y|^2\big)^{-r-1}\big(
g_1(\overrightarrow{\eta},y)+\tau/N\ g_2(\overrightarrow{\eta},y)
+g_3(\overrightarrow{\eta},y)
\big),
\end{equation}
\end{small}
in which
\begin{small}
\begin{equation}\label{g1 eta y}
g_1(\overrightarrow{\eta},y)=\det\left[\begin{smallmatrix}
z\mathbb{I}_r-J\big( \mathbb{I}_r-\overrightarrow{\eta}\overrightarrow{\eta}^*U \big)       &
-\overline{y}U^{-1}  \\
yU  &  
\overline{z}\mathbb{I}_r-J^*\big( \mathbb{I}_r-U\overrightarrow{\eta}\overrightarrow{\eta}^* \big)
\end{smallmatrix}\right],
\end{equation}
\end{small}
\begin{scriptsize}
\begin{multline}\label{g2 eta y}
g_2(\overrightarrow{\eta},y)=
\Big(
(N-r-1)\big(|z|^2+|y|^2\big)^{-1}+2|y|^2\binom{N-r-1}{2}
\big(|z|^2+|y|^2\big)^{-2}
\Big)
\det\left[\begin{smallmatrix}
z\mathbb{I}_r-J\big( \mathbb{I}_r-\overrightarrow{\eta}\overrightarrow{\eta}^*U \big)       &
-\overline{y}U^{-1}  \\
yU  &  
\overline{z}\mathbb{I}_r-J^*\big( \mathbb{I}_r-U\overrightarrow{\eta}\overrightarrow{\eta}^* \big)
\end{smallmatrix}\right]
\\
+\sum_{\alpha,\beta=1}^r 
(-1)^{\alpha+\beta+r}(N-r-1)\big(|z|^2+|y|^2\big)^{-1}
\bigg(
y U_{\alpha,\beta}
\det\left[\begin{smallmatrix}
\big(z\mathbb{I}_r-J\big( \mathbb{I}_r-\overrightarrow{\eta}\overrightarrow{\eta}^*U \big)\big)
\big[ [r];[r]/\{ \beta \} \big]       &
-\overline{y}U^{-1}  \\
yU\big[ [r]/\{ \alpha \};[r]/\{ \beta \} \big]  &  
\big(\overline{z}\mathbb{I}_r-J^*\big( \mathbb{I}_r-U\overrightarrow{\eta}\overrightarrow{\eta}^* \big)\big)
\big[ [r]/\{ \alpha \};[r] \big]
\end{smallmatrix}\right]
\\
-\overline{y} \big(U^{-1}\big)_{\alpha,\beta}
\det\left[\begin{smallmatrix}
\big(z\mathbb{I}_r-J\big( \mathbb{I}_r-\overrightarrow{\eta}\overrightarrow{\eta}^*U \big)\big)
\big[ [r]/\{ \alpha \};[r] \big]       &
-\overline{y}U^{-1}\big[ [r]/\{ \alpha \};[r]/\{ \beta \} \big]  \\
yU  &  
\big(\overline{z}\mathbb{I}_r-J^*\big( \mathbb{I}_r-U\overrightarrow{\eta}\overrightarrow{\eta}^* \big)\big)
\big[ [r];[r]/\{ \beta \} \big]
\end{smallmatrix}\right]\bigg)
+\sum_{\alpha_1,\alpha_2,\beta_1,\beta_2=1}^r
(-1)^{\alpha_1+\alpha_2+\beta_1+\beta_2}
\\
\times U_{\alpha_1,\beta_1}
\big(U^{-1}\big)_{\alpha_2,\beta_2}
\det\left[\begin{smallmatrix}
\big(z\mathbb{I}_r-J\big( \mathbb{I}_r-\overrightarrow{\eta}\overrightarrow{\eta}^*U \big)\big)
\big[ [r]/\{ \alpha_2 \};[r]/\{ \beta_1 \} \big]       &
-\overline{y}U^{-1}\big[ [r]/\{ \alpha_2 \};[r]/\{ \beta_2 \} \big]  \\
yU\big[ [r]/\{ \alpha_1 \};[r]/\{ \beta_1 \} \big]  &  
\big(\overline{z}\mathbb{I}_r-J^*\big( \mathbb{I}_r-U\overrightarrow{\eta}\overrightarrow{\eta}^* \big)\big)
\big[ [r]/\{ \alpha_1 \};[r]/\{ \beta_2 \} \big]
\end{smallmatrix}\right]
\end{multline}
\end{scriptsize}
and
\begin{scriptsize}
\begin{multline}\label{g3 eta y}
g_3(\vec{\eta},y)=
\sum_{\alpha,\beta=1}^r 
(-1)^{\alpha+\beta+r}(N-r-1)\big(|z|^2+|y|^2\big)^{-1}
y 
\big(
J^*U\overrightarrow{\eta}\overrightarrow{\eta}^*UJ
\big( \mathbb{I}_r-\overrightarrow{\eta}\overrightarrow{\eta}^*U \big)
\big)_{\alpha,\beta}
\\
\times 
\det\left[\begin{smallmatrix}
\big(z\mathbb{I}_r-J\big( \mathbb{I}_r-\overrightarrow{\eta}\overrightarrow{\eta}^*U \big)\big)
\big[ [r];[r]/\{ \beta \} \big]       &
-\overline{y}U^{-1}  \\
yU\big[ [r]/\{ \alpha \};[r]/\{ \beta \} \big]  &  
\big(\overline{z}\mathbb{I}_r-J^*\big( \mathbb{I}_r-U\vec{\eta}\vec{\eta}^* \big)\big)
\big[ [r]/\{ \alpha \};[r] \big]
\end{smallmatrix}\right]
+\sum_{\alpha_1,\alpha_2,\beta_1,\beta_2=1}^r
(-1)^{\alpha_1+\alpha_2+\beta_1+\beta_2}\big(U^{-1}\big)_{\alpha_2,\beta_2}
\\
\times \big(
J^*U\overrightarrow{\eta}\overrightarrow{\eta}^*UJ
\big( \mathbb{I}_r-\overrightarrow{\eta}\overrightarrow{\eta}^*U \big)
\big)_{\alpha_1,\beta_1}
\det\left[\begin{smallmatrix}
\big(z\mathbb{I}_r-J\big( \mathbb{I}_r-\overrightarrow{\eta}\overrightarrow{\eta}^*U \big)\big)
\big[ [r]/\{ \alpha_2 \};[r]/\{ \beta_1 \} \big]       &
-\overline{y}U^{-1}\big[ [r]/\{ \alpha_2 \};[r]/\{ \beta_2 \} \big]  \\
yU\big[ [r]/\{ \alpha_1 \};[r]/\{ \beta_1 \} \big]  &  
\big(\overline{z}\mathbb{I}_r-J^*\big( \mathbb{I}_r-U\overrightarrow{\eta}\overrightarrow{\eta}^* \big)\big)
\big[ [r]/\{ \alpha_1 \};[r]/\{ \beta_2 \} \big]
\end{smallmatrix}\right].
\end{multline}
\end{scriptsize}

\end{proposition}

We also need to define a family of  (iterative  erfc) functions, depending on a real parameter  $s>-1$,   
\begin{small}
\begin{equation} \label{IEF}
\mathrm{IE}_{s}(z)= \frac{1}{\sqrt{2\pi} \Gamma(s+1)}
\int_0^{\infty}v^s 
e^{  
-\frac{1}{2}(
v+z)^2}{\rm d}v, \quad z\in \mathbb{C},
\end{equation}
\end{small}
while  as a limit  of    $s>-1$ from  above 
\begin{small}
 \begin{equation} 
\mathrm{IE}_{-1}(z)= \frac{1}{\sqrt{2\pi}  }
e^{  -\frac{1}{2}z^2}.
\end{equation}
\end{small}

For the finite rank perturbation $X_0$, the support of the limit spectral measure $\mu_{\infty}$ of \eqref{model}  is
\begin{equation}
\mathrm{Supp}(\mu_{\infty}):=\Big\{z_0\in \mathbb{C}: 
| z_0 |\leq \sqrt{\tau}
\Big\},
\end{equation}
see e.g. \cite[Corollary 1.12]{TV10} or \cite[Proposition 1.2]{BC16}. Now,
we are ready to  formulate the   main result concerning  scaling limits of  the mean self-overlap at the edge. 
\begin{theorem}\label{OverlapStaEdge}
With the assumptions in \eqref{meanmatrix} and \eqref{Jdetail}, let $r$ be finite. If $|z_0|^2=\tau$, 
 then 
as $N\to \infty$ the scaled mean self-overlap  
\begin{multline}\label{OverlapStaEdgeEqua}
N^{-\frac{1}{2}}
\mathcal{O}_N\big(
\tau,A_0,z_0\big(1+N^{-\frac{1}{2}}\hat{z}\big)
\big)=\sqrt{2/\pi}\Gamma(t+1)/\tau\ 
e^{2
(\Re \hat{z})^2}
\\
\times
\big( (t+1){\rm IE}_{t+1}(
2\Re\hat{z})
{\rm IE}_{t-1}(
-2\Re\hat{z})+t{\rm IE}_{t}(
2\Re\hat{z}){\rm IE}_{t}(
-2\Re \hat{z})
\big)+O\big(
N^{-\frac{1}{4}}
\big),
\end{multline}
uniformly for   all 
$\hat{z} $ in a compact subset of $\mathbb{C}$, where
$t$ is defined in \eqref{geometricmultiplicity}.
\end{theorem}

Next, we consider the cases when $ |z_0|^2>\tau $. Firstly, we consider the case when
\begin{equation}\label{Jdetailoutlier1}
A_0=P R_{r}\left(z_0 \right) P^{-1}.
 \end{equation}
In this case, the perturbed matrix $X$ in definition \ref{GinU} will have outliers located around $z_0$ asymptotically, see e.g. \cite[Theorem 1.8]{BC16}. Here, we give the corresponding asymptotic behaviour for the mean self-overlap at the outlier eigenvalues. Denote $\vec{e}_i$ the standard basis in $\mathbb{R}^r$, for $i=1,\cdots,r$.
\begin{theorem}\label{OverlapStaEdgeOutlier1}
With the assumptions in \eqref{meanmatrix} and \eqref{Jdetailoutlier1}, let $r$ be finite. If $|z_0|^2>\tau$, 
 then 
as $N\to \infty$ the scaled mean self-overlap  
\begin{multline}\label{OverlapStaEdgeEquaOutlier1}
\mathcal{O}_N\big(
\tau,A_0,z_0+N^{-\frac{1}{2r}}\hat{z}
\big)=1/(\pi\tau)\ 
\big(
1-\frac{\tau}{|z_0|^2}
\big)^{-1}
\\
\times
\exp\Big\{-\frac{1}{\tau}
\big(
1-\frac{\tau}{|z_0|^2}
\big)
\big(
\vec{e}_1^*P^*P \vec{e}_1
\big)^{-1}
\big(
\vec{e}_r^*P^{-1}(P^*)^{-1} \vec{e}_r
\big)^{-1}
\big|\hat{z}\big|^{2r}
\Big\}
+O\big(
N^{-\frac{1}{2r}}
\big),
\end{multline}
uniformly for all $\hat{z} $ in a compact subset of $\mathbb{C}$.
\end{theorem}

Secondly, we consider the case when
\begin{equation}\label{Jdetailoutlier2}
A_0=z_0 \mathbb{I}_r.
 \end{equation}
In this case, the perturbed matrix $X$ in definition \ref{GinU} will also have outliers located around $z_0$ asymptotically, see e.g. \cite[Theorem 1.7]{BC16}. Here, we give the corresponding asymptotic behaviour for the mean self-overlap at the corresponding outlier eigenvalues. 
\begin{theorem}\label{OverlapStaEdgeOutlier2}
With the assumptions in \eqref{meanmatrix} and \eqref{Jdetailoutlier2}, let $r$ be finite. If $|z_0|^2>\tau$, 
 then 
as $N\to \infty$ the scaled mean self-overlap  
\begin{multline}\label{OverlapStaEdgeEquaOutlier2}
\mathcal{O}_N\big(
\tau,A_0,z_0+\big(N (\tau^{-1}-|z_0|^{-2})\big)^{-\frac{1}{2}}\hat{z}
\big)=1/(\pi\tau)\ 
\big(
1-\frac{\tau}{|z_0|^2}
\big)^{-1}
\\
\times
\big(
r e_r(|\hat{z}|^2)-|\hat{z}|^2 e_{r-1}(|\hat{z}|^2)
\big)e^{-|\hat{z}|^2}
+O\big(
N^{-\frac{1}{2}}
\big),
\end{multline}
uniformly for all $\hat{z} $ in a compact subset of $\mathbb{C}$, where for any real number $a$, $e_r(a)=\sum_{l=0}^{r-1}\frac{a^l}{l!}$.
\end{theorem}

\begin{remark}\label{RemarkTheoremA}
In Theorem \ref{OverlapStaEdge}, we observe that besides the parameter $\hat{z}$, the asymptotic statistic only depends on a non-negative integer $t$, which corresponds to the geometric multiplicity if the observing point $z_0$ is an eigenvalue of $X_0$, otherwise it is set to  be zero. Based on this observation, Theorem \ref{OverlapStaEdgeOutlier1} and Theorem \ref{OverlapStaEdgeOutlier2}, we get a phenomenon of the phase transition with respect to the modulus of the eigenvalues of $X_0$, that is, extreme eigenvalues(in the sense of modulus) of the deformed  GinUE ensemble    display distinct   patterns in  three different  spectral regimes  created  by the perturbation. For $X_0$ defined in \eqref{meanmatrix}:
\begin{itemize}
\item[(I)] When   all eigenvalues of $X_0$  lie in the open disk  \begin{small}
$ \Big\{z_0\in \mathbb{C}: 
| z_0 |< \sqrt{\tau}
\Big\} $
\end{small}, by setting $t=0$ in \eqref{OverlapStaEdgeEqua}, we get the local statistics for $\mathcal{O}_N\big(\tau,A_0,z\big)$ near the edge point $z_0$:
\begin{equation*}
N^{-\frac{1}{2}}
\mathcal{O}_N\big(
\tau,A_0,z_0\big(1+N^{-\frac{1}{2}}\hat{z}\big)
\big)=\frac{1}{\pi\tau}
\Big(
\frac{1}{\sqrt{2\pi}} e^{-2(\Re \hat{z})^2}
-\Re \hat{z}\times {\rm erfc}
(\sqrt{2}\Re \hat{z})
\Big)
+O\big(
N^{-\frac{1}{4}}
\big),
\end{equation*}
 which coincides with the classical result, see \cite[eqn (29)]{WS2015}, see also 
 \cite[eqn (56)]{GW2018} and \cite[eqn (2.9)]{WCF24}.
 
\item[ (II)]  When  some eigenvalues $\theta_1,\cdots,\theta_g$ of $X_0$ are   on the circle \begin{small}
$ \Big\{z_0\in \mathbb{C}: 
| z_0 |=\sqrt{\tau}
\Big\} $
\end{small} and all others lie   inside it, local statistics for $\mathcal{O}_N\big(\tau,A_0,z\big)$ near the edge point $z_0$ 
will be  characterized by   the iterative erfc  integrals  only when $z_0=\theta_i$ for some $i\in 1,\cdots,g$; Otherwise,   it is the same as the complex Ginibre edge statistcs.

\item[ (III)] When $ X_0={\rm diag}( P R_{r}(z_0 ) P^{-1},0_{N-r} ) $
(resp. $ X_0={\rm diag}( z_0 \mathbb{I}_r,0_{N-r} ) $), and $| z_0 |>\sqrt{\tau}$, outlier eigenvalues of $X$ are located around $z_0$ asymptotically, and local statistics for $\mathcal{O}_N\big(\tau,A_0,z\big)$ near $z_0$ are described by a new kind of function in \eqref{OverlapStaEdgeEquaOutlier1}(resp. \eqref{OverlapStaEdgeEquaOutlier2}).

 \end{itemize}

\end{remark}

\begin{remark}\label{RemarkTheoremB}
In the previous work \cite{LZ12}, under the same model \eqref{model}, a complete phase transition phenomenon have been observed for the statistics near the extreme eigenvalues(in the sense of modulus); While, the journey for completing the phase transition for the eigenvector statistics are far from over. For the model \eqref{model}, one of the most essential issues is the distribution of the overlap, like what have been done in \cite{BD20,Fy18}. This gap is obviously interesting and important, while challenging. 
\end{remark}

 \subsection{Structure of this paper}
 The remainder of  this paper is  organized as follows.   Section \ref{TheoremProofs} is devoted to  the proof of Theorem \ref{OverlapStaEdge}, except that   the proofs of  three propositions are left to Section   \ref{propproofs}.  Section \ref{IntegralRepresentProofs} is devoted to  the proof of Proposition \ref{IntegralRepresent}.

\section{Proof of Theorem \ref{OverlapStaEdge}}\label{TheoremProofs}
In this section we set $|z_0|^2=\tau$ and $\rho=\frac{1}{2}$ in Proposition \ref{IntegralRepresent}, to be more precise, now we consider
\begin{equation}\label{EdgeSetting}
z=z_0+N^{-\frac{1}{2}}\hat{z},\quad
|z_0|^2=\tau.
\end{equation}
 First rewrite
\begin{small}
\begin{equation}\label{logzyrewrite}
\log\big(|z|^2+|y|^2\big)=
\log\big(\tau+|y|^2\big)+
\log\big(
1+(\tau+|y|^2)^{-1}\big(
(z_0\overline{\hat{z}}+\overline{z}_0\hat{z})N^{-\frac{1}{2}}
+N^{-1}|\hat{z}|^2
\big)
\big).
\end{equation}
\end{small}
 Taking all the leading terms of $f(\vec{\eta},y)$ in \eqref{f eta y} and we have
\begin{small}
\begin{equation}\label{f0 eta y}
f_0(\vec{\eta},y)=-\frac{1}{\tau}\vec{\eta}^*
\big( z_0\mathbb{I}_r-J \big)^*U\big( z_0\mathbb{I}_r-J \big)\vec{\eta}
+\log\big( 1-\vec{\eta}^*U\vec{\eta} \big)
+\vec{\eta}^*U\vec{\eta}-\frac{1}{\tau}|y|^2+\log\big(|z|^2+|y|^2\big),
\end{equation}
\end{small}
then by simple analysis we have
\begin{lemma}\label{maximumlemma}
For $\vec{\eta}^*U\vec{\eta}\leq 1$ and any complex number $y$, we have
\begin{small}
\begin{equation}\label{MaximumlemmaEqua}
f_0(\vec{\eta},y)\leq \log(\tau),
\end{equation}
\end{small}
with equality if and only if $\vec{\eta}=\vec{0}$ and $y=0$.
\end{lemma}
Combining the maximum inequality \eqref{MaximumlemmaEqua}, we can restrict the integral representations for the mean overlap statistics given in Proposition \ref{IntegralRepresent} in a small $\delta$-region,
\begin{equation}\label{Omegadelta}
\Omega_{\delta}=\big\{
(\overrightarrow{\eta},y)\big| \overrightarrow{\eta}^*U\overrightarrow{\eta}+|y|^2
\leq \delta
\big\},
\end{equation}
where $\delta>0$ is fixed, independent of $N$, while can be chosen sufficiently small. To be more precise, we have the following concentration reduction
\begin{proposition}\label{foranalysis}
If $z=z_0+N^{-\frac{1}{2}}\hat{z}$ and $|z_0|^2=\tau$, then for any $\delta>0$ which is independent of $N$, there exists $\Delta>0$ such that
\begin{small}
\begin{equation}\label{foranalysisEqua}
\mathcal{O}_N\big(
\tau,A_0,z
\big)=D_N \tau^N\Big(
I_{N,\delta}+O\big(
e^{-\frac{1}{2}N\Delta}
\big)
\Big),
\end{equation}
\end{small}
where
\begin{small}
\begin{equation}\label{INdelta}
I_{N,\delta}=\int_{\Omega_{\delta}} \exp\big\{
N(f(\vec{\eta},y)-\log(\tau))
\big\}g(\vec{\eta},y)
{\rm d}\vec{\eta}{\rm d}y,
\end{equation}
\end{small}
with the same notations as in Proposition \ref{IntegralRepresent}.
\end{proposition}
 
Recalling \eqref{DN}, from the Stirling's formula we have
\begin{small}
\begin{equation}\label{DNasymptotic}
D_N \tau^N=\pi^{-r-2}\tau^{-1}(2\pi N)^{-\frac{1}{2}}N^{r+1}
e^{-\sqrt{N}/\tau\big(
z_0\overline{\hat{z}}+\overline{z}_0\hat{z}
\big)-|\hat{z}|^2/\tau}
\det(U)\big(
1+O(N^{-1})
\big).
\end{equation}
\end{small} 
 The remaining task is to analyze the matrix integral $I_{N,\delta}$, which can be followed by a quite standard procedure of Laplace method.

In view of Proposition \ref{foranalysis}, we will restrict all the matrix variables in the region $\Omega_{\delta}$ defined in \eqref{Omegadelta} for sufficiently small $\delta>0$. Here we must emphasize that $\delta$ is fixed independent of $N$, while we need $\delta>0$ to be sufficiently small so that Taylor expansions of the relevant matrix variables are permitted. As the number of relevant matrix variables is certainly finite,  the required $\delta$ can always be realized. In addition, in the following analysis, all error terms are written as the form $O(A)$, 
which means the error terms are bounded by $CA$ for some positive constant $C$ which is independent of $N$.

First, we have Taylor expansion for $f(\vec{\eta},y)$ on the exponent.
\begin{proposition} \label{fTaylor}
With $f(\vec{\eta},y)$ defined in \eqref{f eta y}, we have 
\begin{small}
\begin{equation}\label{fEtaYexpan}
N\big(
f(\vec{\eta},y)-\log(\tau)
\big)
=F_0+N F_1+O(F_2),
\end{equation}
\end{small}
where
\begin{small}
\begin{equation}\label{F0}
F_0=\sqrt{N}/\tau\big(
\overline{z_0}
\hat{z}
+
z_0
\overline{\hat{z}}
\big)
 +|\hat{z}|^2/\tau,
\end{equation}
\end{small}
\begin{small}
\begin{multline}\label{F1}
F_1=-\overrightarrow{\eta_R}^*\widetilde{U}^{(22)}\overrightarrow{\eta_R}/\tau-
\big(
\overrightarrow{\eta_L}^*U_{I_1}\overrightarrow{\eta_L}
\big)^2/2
\\
+N^{-\frac{1}{2}}/\tau
\big(
\overline{z_0}
\hat{z}+
z_0
\overline{\hat{z}}
\big)
\overrightarrow{\eta_L}^*U_{I_1}\overrightarrow{\eta_L}
-\big(
|y|^2+N^{-\frac{1}{2}}\big(
\overline{z_0}
\hat{z}+
z_0
\overline{\hat{z}}
\big)
\big)^2/(2\tau^2)
\end{multline}
\end{small}
and
\begin{small}
\begin{equation}\label{F2}
F_2=N\| \overrightarrow{\eta} \|^6+
\sqrt{N}\| \overrightarrow{\eta_R} \|\| \overrightarrow{\eta_L} \|
+N\big(
|y|^2+N^{-\frac{1}{2}}
\big)^3.
\end{equation}
$ \widetilde{U}^{(22)} $, $ \overrightarrow{\eta_L} $ and $\overrightarrow{\eta_R} $ are defined in \eqref{Ut22R} and \eqref{etaUcompression1} respectively; $I_{1}$ ($I_{2}$) is the  index set consisting of  numbers corresponding to   first (last)  columns of all 
 blocks  
 $R_{p_{i}}\left(z_0 \right)$(cf. \eqref{1Rpirealedge})    from the Jordan form  $J$ in \eqref{Jdetail}, where  $ 1\leq i \leq m$.
\end{small}

\end{proposition}
We also have Taylor expansion for $g(\vec{\eta},y)$ off the exponent.
\begin{proposition} \label{gTaylor}
With $g(\vec{\eta},y)$ defined in \eqref{g eta y}, we have
\begin{small}
\begin{equation}\label{getayexpan}
g(\vec{\eta},y)=N|y|^{2t}\tau^{-r-1}
\det\big( U[I_1] \big)
\det\big( U^{-1}[I_2] \big)
\big|
\det\big(
z_0\mathbb{I}_{r_0}-J_1
\big)
\big|^2
\big(
|y|^2/\tau+\overrightarrow{\eta_L}^*U_{I_1}\overrightarrow{\eta_L}
\big)+O(\Omega_1),
\end{equation}
\end{small}
where
\begin{small}
\begin{multline}\label{gErrorOmega1}
\Omega_1=N\Big(
|y|^2\big( 
(|y|+N^{-\frac{1}{2}}+\| \overrightarrow{\eta} \|^2)^{2t}-|y|^{2t}
 \big)+|y|\| \overrightarrow{\eta} \|^2
 \big( 
(|y|+N^{-\frac{1}{2}}+\| \overrightarrow{\eta} \|^2)^{2t-1}-|y|^{2t-1}
 \big)
\Big)\\
+N|y|^{2t}\big(
|y|^4+\| \overrightarrow{\eta} \|\| \overrightarrow{\eta_R} \|
\big)+
\big(|y|+N^{-\frac{1}{2}}+\| \overrightarrow{\eta} \|^2\big)^{2t}
+\| \overrightarrow{\eta} \|^2
\big(|y|+N^{-\frac{1}{2}}+\| \overrightarrow{\eta} \|^2 \big)^{2t-2}.
\end{multline}
\end{small}
\end{proposition}

Recalling \eqref{INdelta} and \eqref{fTaylor}, we rewrite
\begin{small}
\begin{equation}\label{INDeltaDecom}
e^{-F_0}I_{N,\delta}=J_{1,N}+J_{2,N},
\end{equation}
\end{small}
where with $\Omega_{\delta}$ defined in \eqref{Omegadelta},
\begin{small}
\begin{equation}\label{J12N}
J_{1,N}=\int_{\Omega_{\delta}}
g\big( \overrightarrow{\eta},y \big)
e^{NF_1}{\rm d}\overrightarrow{\eta}{\rm d}y,
\quad
J_{2,N}=\int_{\Omega_{\delta}}
g\big( \overrightarrow{\eta},y \big)
e^{NF_1}
\big(
e^{O(F_2)}-1
\big)
{\rm d}\overrightarrow{\eta}{\rm d}y.
\end{equation}
\end{small}
Recalling \eqref{F2}, for sufficiently large $N$ and small $\delta$, there exists some $C>0$ such that 
  \begin{small}
\begin{equation*}
\begin{aligned}
&\frac{1}{C} F_2 \leq N^{-\frac{1}{2}}+
\sqrt{N\delta} \big(
\big\| \overrightarrow{\eta_R} \big\|
+|y|^2
\big) +N\delta\big(
\big\| \overrightarrow{\eta_R} \big\|^2
+\big\| \overrightarrow{\eta_L} \big\|^4
+|y|^4
\big).
\end{aligned}
\end{equation*}
\end{small}

 Using the inequality \begin{equation}\label{J2N inequ}
\big|
e^{O(F_2)}-1
\big|\leq O(|F_2|)e^{O(|F_2|)},
\end{equation}
 after change of  variables
\begin{equation}\label{change scale}
\overrightarrow{\eta_R}\rightarrow N^{-\frac{1}{2}}\overrightarrow{\eta_R},\quad
(\overrightarrow{\eta_L},y)\rightarrow 
N^{-\frac{1}{4}}(\overrightarrow{\eta_L},y),
\end{equation}
the term $O(F_2)$ in \eqref{J2N inequ} has an upper bound  by   
$N^{-\frac{1}{4}}P\big( \overrightarrow{\eta_L},\overrightarrow{\eta_R},y \big)$ for some polynomial of variables.  
since $F_1$ can control $F_2$  for sufficiently  small $\delta$, by the argument of  Laplace method and the dominant convergence theorem we know that 
$J_{2,N}$ is typically of order  $N^{-\frac{1}{4}}$ compared with $J_{1,N}$, that is,  
 \begin{equation}\label{J2Nestimation}
J_{2,N}=O\big(
N^{-\frac{1}{4}}
\big)J_{1,N}.
\end{equation}
For $J_{1,N}$,   
take  a large  $M_0>0$ such that
\begin{small}
\begin{equation*}
\Omega_{\delta}^{\complement}\subseteq   
\Big\{
|y|^2>\frac{\delta}{M_0}
\Big\}
\bigcup
\Big\{
\overrightarrow{\eta_L}^*\overrightarrow{\eta_L}
>\frac{\delta}{M_0}
\Big\} \bigcup \Big\{
\overrightarrow{\eta_R}^*\overrightarrow{\eta_R}
>\frac{\delta}{M_0}
\Big\}
\end{equation*}
\end{small}
For each piece of  domain,  only  keep  the restricted variable and let  the others free,  it's easy to prove that the corresponding  matrix integral  is exponentially  small, that is,  $O\big( e^{-\delta_1 N} \big)
$
for some $\delta_1>0$.  

So we can extend the integration  region from $\Omega_{\delta}$ to the whole integration region. Therefore, combining \eqref{INdelta}, \eqref{fEtaYexpan}, \eqref{getayexpan}, \eqref{INDeltaDecom}, \eqref{J12N} and \eqref{J2Nestimation}, after the rescaling \eqref{change scale}, we arrive at the following matrix integral
\begin{small}
\begin{equation}\label{INdeltaDecomF}
e^{-F_0}I_{N,\delta}=
\tau^{-r-1}\det\big( U[I_1] \big)
\det\big( U^{-1}[I_2] \big)
\big| \det\big( z_0\mathbb{I}_{r_0}-J_1 \big) \big|^2
N^{-r}\big(
I_0+O\big(
N^{-\frac{1}{4}}
\big)
\big),
\end{equation}
\end{small}
where
\begin{small}
\begin{multline}\label{I0}
I_0=\int |y|^{2t}\big(
|y|^2/\tau+\overrightarrow{\eta_L}^*U_{I_1}\overrightarrow{\eta_L}
\big)\exp\Big\{
-\overrightarrow{\eta_R}^*\widetilde{U}^{(22)}\overrightarrow{\eta_R}/\tau
\\
-\big( \overrightarrow{\eta_L}^*U_{I_1}\overrightarrow{\eta_L} \big)^2/2
+\big( z_0\overline{\hat{z}}+\overline{z}_0\hat{z} \big)
\overrightarrow{\eta_L}^*U_{I_1}\overrightarrow{\eta_L}/\tau
-\big( |y|^2+z_0\overline{\hat{z}}+\overline{z}_0\hat{z} \big)^2
/(2\tau^2)
\Big\}{\rm d}\overrightarrow{\eta_R}{\rm d}\overrightarrow{\eta_L}
{\rm d}y.
\end{multline}
\end{small}
Note the integral with respect to $\overrightarrow{\eta_R}$ is in fact a Gaussian integral
\begin{small}
\begin{equation}\label{GaussEtaR}
\int e^{-\overrightarrow{\eta_R}^*\widetilde{U}^{(22)}\overrightarrow{\eta_R}/\tau}
{\rm d}\overrightarrow{\eta_R}=(\pi\tau)^{r-t}\big(
\det\big( \widetilde{U}^{(22)} \big)
\big)^{-1}.
\end{equation}
\end{small}
For the integration variables $\overrightarrow{\eta_L}$ and $y$, first apply polar coordinate transformation
\begin{small}
\begin{equation*}
y=\sqrt{r}e^{{\rm i}\theta},\quad
{\rm d}y={\rm d}{\rm d}\theta/2.
\end{equation*}
\end{small}
Obviously, the integrand in $I_0$ is independent from $\theta$, integrating out $\theta$,  making a rescaling
\begin{small}
\begin{equation*}
\overrightarrow{\eta_L}\rightarrow \big( U[I_1] \big)^{-\frac{1}{2}}\overrightarrow{\eta_L}, \quad
r\rightarrow \tau r,
\end{equation*}
\end{small}
and further applying sphere coordinate transformation 
\begin{small}
\begin{equation*}
\overrightarrow{\eta_L}=\sqrt{v}\overrightarrow{u_L}, \quad
\overrightarrow{u_L}^*\overrightarrow{u_L}=1,\quad
{\rm d}\overrightarrow{\eta_L}=\frac{\pi^t}{\Gamma(t)}
v^{t-1}{\rm d}v{\rm d}\overrightarrow{u_L},
\end{equation*}
\end{small}
with $\overrightarrow{u_L}$ chosen from $\mathbb{S}^{2t-1}(1,\mathbb{C})$(the surface of the unit sphere in $\mathbb{C}^t$) equipped with the Haar probability measure. After basic computations we have
\begin{small}
\begin{multline}\label{I0Final}
I_0=\frac{(\pi \tau)^{r+1}}{\Gamma(t)}\Big(
\det\big( \widetilde{U}^{(22)} \big)
\det\big( U[I_1] \big)
\Big)^{-1}
\\
\times
\int_0^{+\infty}\int_0^{+\infty}
r^t(r+v)v^{t-1}
e^{-v^2/2+2\Re (z_0^{-1}\hat{z})v
-\big( r+2\Re (z_0^{-1}\hat{z}) \big)^2/2}
{\rm d}v{\rm d}r.
\end{multline}
\end{small}

By applying the Schur complement technique we have the following identity
\begin{small}
\begin{equation}\label{SchurIden}
\det\big(U\big)
\big(\det\big(\widetilde{U}^{(22)}\big)\big)^{-1}
\det\big(U^{-1}[I_2]\big)
\big|\det\big(z_0\mathbb{I}_{r_0}-J_1\big)\big|^2
=1.
\end{equation}
\end{small} 

Combining \eqref{foranalysisEqua}, \eqref{F0}, \eqref{DNasymptotic}, \eqref{INdeltaDecomF}, \eqref{I0Final} and \eqref{SchurIden}, we finally arrive at
\begin{small}
\begin{multline*}
N^{-\frac{1}{2}}\mathcal{O}_N(\tau,A_0,z)=
\big( \sqrt{2\pi}\Gamma(t)\pi\tau \big)^{-1}
\int_0^{+\infty}\int_0^{+\infty}
r^t(r+v)v^{t-1}
\\
\times
e^{-v^2/2+2\Re (z_0^{-1}\hat{z})v
-\big( r+2\Re (z_0^{-1}\hat{z}) \big)^2/2}
{\rm d}v{\rm d}r+O\big(N^{-\frac{1}{4}}\big).
\end{multline*}
\end{small}

\section{Proof of Proposition \ref{IntegralRepresent}.}\label{IntegralRepresentProofs}
{\bf Step 1: Partial Schur decomposition.} With probability 1 we can assume the eigenvalues of $X$, $z_1,\cdots,z_N$ to be pairwise distinct. Based on this observation, using the lexicographic order on complex numbers, that is
\begin{small}
\begin{equation*}
u+{\rm i}v\leq u'+{\rm i}v'\ \ 
\textit{if}\ \ u<u'\ \ \textit{or}\ \ \textit{if}\ \ u=u'
\ \ \textit{and}\ \ v\leq v',
\end{equation*}
\end{small}
we can arrange the eigenvalues in increasing order. Thus,
\begin{small}
\begin{equation}\label{eigenvalueorder}
z_1<\cdots<z_N.
\end{equation}
\end{small}

Further in \eqref{OverlapStatistic}, for the n-th summation, we can apply the partial Schur decomposition
\begin{small}
\begin{equation}\label{PartialSchur}
X_N=U\widetilde{X}_N U^*,\quad
\widetilde{X}_N=\begin{bmatrix}
z_n & \overrightarrow{\omega}^*
\\ & X_{N-1}
\end{bmatrix},
\end{equation}
\end{small}
here $U=[\overrightarrow{u_1},U_2(\overrightarrow{u_1})]$,
\begin{small}
\begin{equation}\label{PartialSchurU}
\overrightarrow{u_1}\in \mathcal{B}_N,\quad
\mathcal{B}_N=\big\{
\overrightarrow{u}\in \mathbb{C}^{N}\big|
\overrightarrow{u}^*\overrightarrow{u}=1,\ \ 
\big(\overrightarrow{u}\big)_N\geq 0.
\big\}.
\end{equation}
\end{small}
Define the eigenvalues of $X_{N-1}$ to be \begin{small}
$\big\{
z_1(X_{N-1})<
\cdots<z_{N-1}(X_{N-1})
\big\},$
\end{small} in an increasing order. This can be done as the eigenvalues set of $X_N$ is the same as that of $\widetilde{X}_N$. From \eqref{eigenvalueorder} we know
 $\big(z_n,X_{N-1}\big)\in \mathcal{A}_n$, where $\mathcal{A}_n$ is a subset of
  $\mathbb{C}\times \mathbb{C}^{(N-1)\times (N-1)}$ defined by
\begin{small}
\begin{equation}\label{mathcal An}
\mathcal{A}_n=\big\{
\big(z_n,X_{N-1}\big)\big|
z_1(X_{N-1})<\cdots<z_{n-1}(X_{N-1})
<z_n<z_{n}(X_{N-1})<\cdots<z_{N-1}(X_{N-1})
\big\}.
\end{equation}
\end{small}
Obviously, 
\begin{small}
\begin{equation}\label{AnCup}
\bigcup_{n=1}^N \mathcal{A}_n=\mathbb{C}
\times \mathbb{C}^{(N-1)\times (N-1)}.
\end{equation}
\end{small}
 By requiring
\begin{small}
\begin{equation}\label{DecomUnique}
\big(
\overrightarrow{\omega},z_n,X_{N-1},U
\big)\in
 \mathbb{C}^{N-1}\times \mathcal{A}_n \times \mathcal{B}_n,
\end{equation}
\end{small}
 we can restore uniqueness of the decomposition \eqref{PartialSchurU}. Moreover, we have the Jacobian of the decomposition \eqref{PartialSchurU}:
 \begin{small}
 \begin{equation}\label{PSJacobian}
 {\rm d}X_N=\det(\Sigma(z_n))
 {\rm d}z_n{\rm d}\overrightarrow{\omega}
 {\rm d}S(\overrightarrow{u_1})
 {\rm d}X_{N-1},
 \end{equation}
 \end{small}
 where ${\rm d}S(\overrightarrow{u_1})$ is the volume element of the surface of the unit sphere, and
 \begin{small}
 \begin{equation}\label{Sigmaz}
 \Sigma(z)=(z\mathbb{I}_{N-1}-X_{N-1})^*(z\mathbb{I}_{N-1}-X_{N-1}).
 \end{equation}
 \end{small}
 We provide a proof of \eqref{PSJacobian} in Appendix \ref{Appendix1}, 

Note that we can represent $\mathcal{O}_{n,n}$ in terms of $(z_n,\overrightarrow{\omega},X_{N-1})$. To be more specifically, firstly, from \eqref{PartialSchur} we know the n-th self overlap statistic of $\widetilde{X}_{N-1}$
is the same as that of $X_{N-1}$. Again, in view of \eqref{PartialSchur} from the structure of $\widetilde{X}_{N-1}$, we know $\widetilde{X}_{N-1}$ has the right eigenvector
 $\widetilde{v}_R=(1,\overrightarrow{0}_{_{N-1}}^{\rm T})^{\rm T},$ by the bi-orthogonality, we may set $\widetilde{v}_L=(1,\overrightarrow{b}_{_{N-1}}^{\rm T})^{\rm T}$, by solving the equation
  $\widetilde{v}_L^*\widetilde{X}_{N-1}=z_n\widetilde{v}_L^*, $ we get that $\overrightarrow{b}_{N-1}^{^*}=\overrightarrow{\omega}^*(z_n\mathbb{I}_{N-1}-X_{N-1})^{-1}$. Hence, 
  \begin{small}
  \begin{equation}\label{nthOverlap}
  \mathcal{O}_{n,n}=1+\overrightarrow{\omega}^*
  \Sigma(z_n)^{-1}
  \overrightarrow{\omega}.
  \end{equation}
  \end{small}
  
{\bf Step 2: Multiple integrals.} Now, starting from the procedure in {\bf Step 1}, in \eqref{model},\begin{small}
\begin{equation*}
-(N/\tau){\rm Tr}(X_N-X_0)(X_N-X_0)^*
=-(N/\tau){\rm Tr}(\widetilde{X}_N-B)(\widetilde{X}_N-B)^*,
\end{equation*}
\end{small}
\begin{small}
where
\begin{equation}\label{B}
B=U^*X_0U:=\begin{bmatrix}
b_{1,1} & B_{1,2}^*  \\
B_{2,1} & B_{2,2}
\end{bmatrix},
\end{equation}
\end{small}
from $U=[\overrightarrow{u_1},U_2=U_2(\overrightarrow{u_1})]$ we know 
\begin{small}
\begin{equation}\label{Bdetail}
b_{1,1}=\overrightarrow{u_1}^*X_0\overrightarrow{u_1},\quad
B_{1,2}^*=\overrightarrow{u_1}^*X_0U_2,\quad
B_{2,1}=U_2^*X_0\overrightarrow{u_1},\quad
B_{2,2}=U_2^*X_0U_2.
\end{equation}
\end{small}
As $X_0$ admits the block structure \eqref{meanmatrix}, we also need to develop the same block structure for $\overrightarrow{u_1}$ and $U_2$,
\begin{small}
\begin{equation}\label{Ublockstructure}
\overrightarrow{u_1}=\left[\begin{smallmatrix}
\overrightarrow{q} \\
\widetilde{u}_1
\end{smallmatrix}\right],\quad
U_2=\left[\begin{smallmatrix}
U_{1,2}  \\ U_{2,2}
\end{smallmatrix}\right],
\end{equation}
\end{small}
where $\overrightarrow{q}$ and $\widetilde{u}_1$ are column vectors of sizes $r$ and $N-r$ respectively; $U_{1,2}$ and $U_{2,2}$ are of sizes $r\times (N-1)$ and $(N-r)\times (N-1)$ respectively. Combining \eqref{meanmatrix} and \eqref{Ublockstructure} we have
\begin{small}
\begin{equation}\label{BdetailF}
b_{1,1}=\overrightarrow{q}^*A_0\overrightarrow{q},\quad
B_{1,2}^*=\overrightarrow{q}^*A_0U_{1,2},\quad
B_{2,1}=U_{1,2}^*A_0\overrightarrow{q},\quad
B_{2,2}=U_{1,2}^*A_0U_{1,2}.
\end{equation}
\end{small}
From the identity $\overrightarrow{q}\overrightarrow{q}^*+U_{1,2}U_{1,2}^*=\mathbb{I}_r$, by singular value decomposition we have the decomposition
\begin{small}
\begin{equation}\label{U12SVD}
U_{1,2}=\begin{bmatrix}
\sqrt{\mathbb{I}_r-\overrightarrow{q}\overrightarrow{q}^*}
& 0_{r\times(N-r-1)}
\end{bmatrix}V,\quad
V\in \mathcal{U}(N-1).
\end{equation}
\end{small}

Now we can represent the n-th summation in \eqref{OverlapStatistic} as a multiple integral
\begin{small}
\begin{multline}\label{OverMultiple}
\mathbb{E}\big(
\mathcal{O}_{n,n}\delta(z-z_n)
\big)=\Big(
\frac{N}{\pi\tau}
\Big)^{N^2}
\int_{\mathcal{A}_n}\int_{\mathbb{C}^{N-1}}\int_{\mathcal{B}_N}
\big( 1+\overrightarrow{\omega}\Sigma(z_n)^{-1}\overrightarrow{\omega} \big)
\det(\Sigma(z_n))e^{-(N/\tau)B_{2,1}^*B_{2,1}}
\\
\times
e^{-\frac{N}{\tau}|z_n-b_{1,1}|^2-\frac{N}{\tau}
(\overrightarrow{\omega}-B_{1,2})^*
(\overrightarrow{\omega}-B_{1,2})
-\frac{N}{\tau}{\rm Tr}(X_{N-1}-B_{2,2})(X_{N-1}-B_{2,2})^*}\delta(z-z_n){\rm d}z_n{\rm d}X_{N-1}
{\rm d}\overrightarrow{\omega}{\rm d}S(\overrightarrow{u_1}).
\end{multline}
\end{small}

 By multiplying an extra factor $1/(2\pi)$ outside the integral, we can extend $\mathcal{B}_N$(cf. \eqref{PartialSchurU}) to the whole surface of the unit sphere in $\mathbb{C}^N$; Also we can replace the volume element ${\rm d}S(\overrightarrow{u_1})$ with 
\newline
 $\delta(1-\sqrt{\overrightarrow{u_1}^*\overrightarrow{u_1}})=2\delta(1-\overrightarrow{u_1}^*\overrightarrow{u_1})$; and from \eqref{AnCup}, by summing $n$ from $1$ to $N$, in \eqref{OverMultiple} we can replace the integration region $\mathcal{A}_n$ with the whole space $\mathbb{C}\times \mathbb{C}^{(N-1)\times (N-1)}$, and use a unified variable $\widetilde{z}$ to replace $z_n$ for each $n$; After integrating $\widetilde{z}$ with help of $\delta(z-\widetilde{z})$, we arrive at
\begin{small}
\begin{multline}\label{OverMultiple1}
\mathcal{O}_N\big(
\tau,A_0,z
\big)=
\frac{1}{N}
\sum_{n=1}^N\mathbb{E}\Big(
 \mathcal{O}_{n,n}\delta(z-z_n)
\Big)=\frac{1}{N\pi}\Big(
\frac{N}{\pi}
\Big)^{N^2}
\\
\times
\int_{ \mathbb{C}^{(N-1)\times (N-1)}}
\int_{\mathbb{C}^{N-1}}\int_{\mathbb{C}_N}
\big( 1+\overrightarrow{\omega}\Sigma(z)^{-1}\overrightarrow{\omega} \big)
\det(\Sigma(z))\delta(1-\overrightarrow{u_1}^*\overrightarrow{u_1})
e^{-(N/\tau)B_{2,1}^*B_{2,1}}
\\
\times
e^{-\frac{N}{\tau}|z-b_{1,1}|^2-\frac{N}{\tau}
(\overrightarrow{\omega}-B_{1,2})^*
(\overrightarrow{\omega}-B_{1,2})
-\frac{N}{\tau}{\rm Tr}(X_{N-1}-B_{2,2})(X_{N-1}-B_{2,2})^*}
{\rm d}X_{N-1}
{\rm d}\overrightarrow{\omega}{\rm d}\overrightarrow{u_1}.
\end{multline}
\end{small}

To integrate out $\overrightarrow{\omega}$, first consider
\begin{small}
\begin{equation*}
f(t)=\int_{\mathbb{C}^{N-1}}
e^{t(1+\overrightarrow{\omega}^*\Sigma(z)^{-1}\overrightarrow{\omega})
-\frac{N}{\tau}
(\overrightarrow{\omega}-B_{1,2})^*
(\overrightarrow{\omega}-B_{1,2})}
{\rm d}\overrightarrow{\omega},
\end{equation*}
\end{small}
which is indeed a Gaussian integral with respect to $\overrightarrow{\omega}$ and equals to
\begin{small}
\begin{equation*}
\big(
\pi\tau/N
\big)^{N-1}
\big(\det\big(
\mathbb{I}_{N-1}-(t\tau/N)\Sigma(z)^{-1}
\big)\big)^{-1}
e^{
t+\frac{N}{\tau}B_{1,2}^*\big(
\mathbb{I}_{N-1}-(t\tau/N)\Sigma(z)^{-1}
\big)^{-1}-\frac{N}{\tau}B_{1,2}^*B_{1,2}
}.
\end{equation*}
\end{small}
By elementary series expansions we can get that
\begin{small}
\begin{multline}\label{omegaIntegral}
\int_{\mathbb{C}^{N-1}}
\big(1+\overrightarrow{\omega}^*\Sigma(z)^{-1}\overrightarrow{\omega}\big)
e^{
-\frac{N}{\tau}
(\overrightarrow{\omega}-B_{1,2})^*
(\overrightarrow{\omega}-B_{1,2})}
{\rm d}\overrightarrow{\omega}
\\
=\big(
{\rm d}f(t)/{\rm d}t
\big)\Big|_{t=0}=\big(
\pi\tau/N
\big)^{N-1}
\big(
1+(\tau/N){\rm Tr}\big(
\Sigma(z)^{-1}
\big)+B_{1,2}^*\Sigma(z)^{-1}B_{1,2}
\big).
\end{multline}
\end{small}

{\bf Step 3: Determinantal identity.} By differentiating the identity
\begin{small}
\begin{equation*}
\det\big(
\mu B_{1,2}B_{1,2}^*+\Sigma
\big)=\det(\Sigma)
e^{{\rm Tr}\log\big(
\mathbb{I}_{N-1}+\mu B_{1,2}B_{1,2}^*\Sigma^{-1}\big)
}=\det(\Sigma) e^{
\sum_{k=1}^{+\infty}\frac{(-1)^{k-1}}{k}
\mu^k {\rm Tr}( B_{1,2}B_{1,2}^*\Sigma^{-1} )^k
},
\end{equation*}
\end{small}
and using \eqref{Sigmaz} we can get that
\begin{small}
\begin{equation}\label{detTdent1}
\det\big(
\Sigma(z)
\big)\big(
B_{1,2}\Sigma(z)^{-1}B_{1,2}^*
\big)=\frac{\partial}{\partial \mu}
\det\begin{bmatrix}
z\mathbb{I}_{N-1}-X_{N-1} & -\sqrt{\mu}\mathbb{I}_{N-1}
\\
\sqrt{\mu}B_{1,2}B_{1,2}^* & \overline{z}\mathbb{I}_{N-1}-X_{N-1}^*
\end{bmatrix}\bigg|_{\mu=0}.
\end{equation}
\end{small}
In the same spirit, we have
\begin{small}
\begin{equation}\label{detTdent2}
\det\big(
\Sigma(z)
\big){\rm Tr}\big(
\Sigma(z)^{-1}
\big)=\frac{\partial}{\partial \mu}
\det\begin{bmatrix}
z\mathbb{I}_{N-1}-X_{N-1} & -\sqrt{\mu}\mathbb{I}_{N-1}
\\
\sqrt{\mu}\mathbb{I}_{N-1} & \overline{z}\mathbb{I}_{N-1}-X_{N-1}^*
\end{bmatrix}\bigg|_{\mu=0}.
\end{equation}
\end{small}
Also from \eqref{Sigmaz} we can rewrite
\begin{small}
\begin{equation}\label{detTdent3}
\det\big(
\Sigma(z)
\big)=
\det\begin{bmatrix}
z\mathbb{I}_{N-1}-X_{N-1} &
\\
 & \overline{z}\mathbb{I}_{N-1}-X_{N-1}^*
\end{bmatrix}.
\end{equation}
\end{small}

As for the integration over $X_{N-1}$, we can apply the following duality formula
\begin{small}
\begin{multline}\label{dualityFormula}
\int_{\mathbb{C}^{(N-1)\times(N-1)}}
e^{-\frac{N}{\tau}{\rm Tr}
(X_{N-1}-F)(X_{N-1}-F)^*
}
\det\begin{bmatrix}
z\mathbb{I}_{N-1}-X_{N-1} & D
\\
E & \overline{z}\mathbb{I}_{N-1}-X_{N-1}^*
\end{bmatrix}{\rm d}X_{N-1}
\\
=\Big(
\frac{\pi\tau}{N}
\Big)^{N^2-2N}\int_{\mathbb{C}}
e^{-\frac{N}{\tau}}\det\begin{bmatrix}
z\mathbb{I}_{N-1}-F & D-\overline{y}\mathbb{I}_{N-1}
\\
E+y\overline{I}_{N-1} & \overline{z}\mathbb{I}_{N-1}-F*
\end{bmatrix}{\rm d}y,
\end{multline}
\end{small}
By applying \eqref{dualityFormula}, we can convert the $2(N-1)^2$ dimensional integral to a integral with dimension 2. Now combining \eqref{OverMultiple1}, \eqref{omegaIntegral}, \eqref{detTdent1}-\eqref{dualityFormula}, we can get an intermediate result
\begin{lemma}\label{intermediatelemma}
With $b_{1,1}$, $B_{1,2}$, $B_{2,1}$ and $B_{2,2}$ given in \eqref{BdetailF}, we have
\begin{small}
\begin{equation}\label{intermediatelemmaEqua}
\mathcal{O}_N\big(
\tau,A_0,z
\big)=(1/(N\pi))(N/(\pi\tau))^{N+1}
I(\mu)|_{\mu=0},
\end{equation}
\end{small}
where
\begin{small}
\begin{multline}\label{ImuI}
I(\mu)=\int
e^{-\frac{N}{\tau}|z-b_{1,1}|^2
-\frac{N}{\tau}B_{2,1}^*B_{2,1}-\frac{N}{\tau}|y|^2
}
\delta(1-\overrightarrow{u_1}^*\overrightarrow{u_1})
\\
\times \bigg(
\det\begin{bmatrix}
z\mathbb{I}_{N-1}-B_{2,2} & -\overline{y}\mathbb{I}_{N-1}
\\
y\mathbb{I}_{N-1} & \overline{z}\mathbb{I}_{N-1}-B_{2,2}^*
\end{bmatrix}+\frac{\tau}{N}
\frac{\partial}{\partial \mu}\det\begin{bmatrix}
z\mathbb{I}_{N-1}-B_{2,2} & -(\sqrt{\mu}+\overline{y})\mathbb{I}_{N-1}
\\
(\sqrt{\mu}+y)\mathbb{I}_{N-1} & \overline{z}\mathbb{I}_{N-1}-B_{2,2}^*
\end{bmatrix}
\\
+\frac{\partial}{\partial \mu}\det\begin{bmatrix}
z\mathbb{I}_{N-1}-B_{2,2} & -(\sqrt{\mu}+\overline{y})\mathbb{I}_{N-1}
\\
\sqrt{\mu}B_{1,2}B_{1,2}^*+y\mathbb{I}_{N-1} & \overline{z}\mathbb{I}_{N-1}-B_{2,2}^*
\end{bmatrix}\bigg){\rm d}\overrightarrow{u_1}{\rm d}y.
\end{multline}
\end{small}
\end{lemma}

{\bf Step 5.} \begin{proof}[Poof of Proposition \ref{IntegralRepresent}.]
Now we start from Lemma \ref{intermediatelemma}. Recalling $z=z_0+N^{-\rho}\hat{z}$ and \eqref{BdetailF}, from basic computations we have
\begin{small}
\begin{multline}\label{exponentIntegral}
-(N/\tau)|z-b_{1,1}|^2-(N/\tau)B_{2,1}^*B_{2,1}=
-(N/\tau)\big|z_0+N^{-\rho}\hat{z}\big|^2
\\
+(N^{1-\rho}/\tau)\big(
\hat{z}\overrightarrow{q}^*A_0^*\overrightarrow{q}
+
\overline{\hat{z}}\overrightarrow{q}^*A_0\overrightarrow{q}
\big)
-(N/\tau)\overrightarrow{q}^*\big(
z_0\mathbb{I}_{r_0}-A_0
\big)^*\big(
z_0\mathbb{I}_{r_0}-A_0
\big)\overrightarrow{q}+(N|z_0|^2/\tau)
\overrightarrow{q}^*\overrightarrow{q}.
\end{multline}
\end{small}

From \eqref{Ublockstructure} we can rewrite $\delta(1-\overrightarrow{u_1}^*\overrightarrow{u_1})=\delta(1-\overrightarrow{q}^*\overrightarrow{q}-\widetilde{u}_1^*\widetilde{u}_1)$, and set $\widetilde{u}_1=\widehat{u}_1\sqrt{1-\overrightarrow{q}^*\overrightarrow{q}}$, we have
\begin{small}
\begin{equation}\label{Volume times Dirac}
\delta(1-\overrightarrow{u_1}^*\overrightarrow{u_1}){\rm d}\overrightarrow{u_1}
=
\big(
1-\overrightarrow{q}^*\overrightarrow{q}
\big)^{N-r-1}\delta(1-\widehat{u}_1\widehat{u}_1)
{\rm d}\overrightarrow{q}
{\rm d}\widehat{u}_1,
\end{equation}
\end{small}
with
\begin{small}
\begin{equation}\label{qvec Restriction}
0\leq \overrightarrow{q}^*\overrightarrow{q} \leq 1.
\end{equation}
\end{small}
Noting from \eqref{exponentIntegral} and \eqref{BdetailF}, we know the terms on the exponential, $b_{1,1}$, $B_{1,2}$, $B_{2,1}$ and $B_{2,2}$ are all independent from $\widehat{u}_1$. So we can integrate $\widehat{u}_1$ out by
\begin{small}
\begin{equation}\label{hatu1Integration}
\int_{\mathbb{C}^{N-r}} \delta(1-\widehat{u}_1\widehat{u}_1){\rm d}\widehat{u}_1
=\pi^{N-r}/(N-r-1)!.
\end{equation}
\end{small}

From \eqref{BdetailF} and \eqref{U12SVD}, by performing elementary operations for matrices with the help of block structure of $U_{1,2}$, we can cancel out the unitary matrix $V$; And by multiplying \begin{small}
${\rm diag}\big(
\big(\mathbb{I}_{r}-\overrightarrow{q}\overrightarrow{q}^*\big)^{-\frac{1}{2}}
,
\big(\mathbb{I}_{r}-\overrightarrow{q}\overrightarrow{q}^*\big)^{-\frac{1}{2}}
\big)$ \end{small} on the left side, and multiplying \begin{small}
${\rm diag}\big(
\big(\mathbb{I}_{r}-\overrightarrow{q}\overrightarrow{q}^*\big)^{\frac{1}{2}}
,
\big(\mathbb{I}_{r}-\overrightarrow{q}\overrightarrow{q}^*\big)^{\frac{1}{2}}
\big)$\end{small}  on the right side, we can have
\begin{small}
\begin{equation}\label{determinantSimplyfy1}
\det\left[\begin{smallmatrix}
z\mathbb{I}_{N-1}-B_{2,2} & -(\sqrt{\mu}+\overline{y})\mathbb{I}_{N-1}
\\
(\sqrt{\mu}+y)\mathbb{I}_{N-1} & \overline{z}\mathbb{I}_{N-1}-B_{2,2}^*
\end{smallmatrix}\right]=
\det\left[\begin{smallmatrix}
z\mathbb{I}_{r}-A_0\big( \mathbb{I}_r-\overrightarrow{q}\overrightarrow{q}^* \big) & 
-(\sqrt{\mu}+\overline{y})\mathbb{I}_{r}
\\
(\sqrt{\mu}+y)\mathbb{I}_{r} & 
\overline{z}\mathbb{I}_{r}-A_0^*\big( \mathbb{I}_r-\overrightarrow{q}\overrightarrow{q}^* \big)
\end{smallmatrix}\right]\big(
|z|^2+|\sqrt{\mu}+y|^2
\big)^{N-r-1},
\end{equation}
\end{small}
and
\begin{small}
\begin{multline}\label{determinantSimplyfy2}
\det\left[\begin{smallmatrix}
z\mathbb{I}_{N-1}-B_{2,2} & -(\sqrt{\mu}+\overline{y})\mathbb{I}_{N-1}
\\
\sqrt{\mu}B_{1,2}B_{1,2}^*+y\mathbb{I}_{N-1} & \overline{z}\mathbb{I}_{N-1}-B_{2,2}^*
\end{smallmatrix}\right]
\\
=
\det\left[\begin{smallmatrix}
z\mathbb{I}_{r}-A_0\big( \mathbb{I}_r-\overrightarrow{q}\overrightarrow{q}^* \big) & 
-(\sqrt{\mu}+\overline{y})\mathbb{I}_{r}
\\
\sqrt{\mu}A_0^*
\overrightarrow{q}\overrightarrow{q}^*A_0\big( \mathbb{I}_r-\overrightarrow{q}\overrightarrow{q}^* \big)
+y\mathbb{I}_{r} & 
\overline{z}\mathbb{I}_{r}-A_0^*\big( \mathbb{I}_r-\overrightarrow{q}\overrightarrow{q}^* \big)
\end{smallmatrix}\right]\big(
|z|^2+y(\sqrt{\mu}+\overline{y})
\big)^{N-r-1},
\end{multline}
\end{small}

Now change the variable
\begin{small}
\begin{equation}\label{changevariable}
\overrightarrow{q}=P\overrightarrow{\eta},
\end{equation}
\end{small}
with $U=P^*P$(cf. \eqref{U}), we have the Jacobian
\begin{small}
\begin{equation}\label{qetaJacobian}
{\rm d}\overrightarrow{q}=\det(U){\rm d}\overrightarrow{\eta}.
\end{equation}
\end{small}

From \eqref{Jdetail}, in \eqref{exponentIntegral}, we have
\begin{small}
\begin{equation}\label{exponentqetaS}
\overrightarrow{q}^*A_0\overrightarrow{q}=
\overrightarrow{\eta}^*UJ\overrightarrow{\eta},\quad
\overrightarrow{q}^*
\big(
z_0\mathbb{I}_r-A_0
\big)^*
\overrightarrow{q}^*
\big(
z_0\mathbb{I}_r-A_0
\big)
\overrightarrow{q}=
\overrightarrow{\eta}^*
\big(
z_0\mathbb{I}_r-J
\big)^*U
\overrightarrow{q}^*
\big(
z_0\mathbb{I}_r-J
\big)
J\overrightarrow{\eta}.
\end{equation}
\end{small} 
Also in \eqref{determinantSimplyfy1} and \eqref{determinantSimplyfy2}, we have
 \begin{small}
 \begin{multline}\label{deterqetaS}
 A_0\big( \mathbb{I}_r-\overrightarrow{q}\overrightarrow{q}^* \big)
 =PJ
 \big( \mathbb{I}_r
 -\overrightarrow{\eta}\overrightarrow{\eta}^*U \big)P^{-1},\quad
  A_0^*\big( \mathbb{I}_r-\overrightarrow{q}\overrightarrow{q}^* \big)
 =(P^{-1})^*J^*
 \big( \mathbb{I}_r
 -U\overrightarrow{\eta}\overrightarrow{\eta}^* \big)P^*,
 \\
A_0^*\overrightarrow{q}\overrightarrow{q}^*A_0\big( \mathbb{I}_r-\overrightarrow{q}\overrightarrow{q}^* \big)=
(P^{-1})^*J^*
U\overrightarrow{\eta}\overrightarrow{\eta}^*
UJ
 \big( \mathbb{I}_r
 -\overrightarrow{\eta}\overrightarrow{\eta}^*U \big)P^{-1}.
 \end{multline}
 \end{small}
 Hence, 
  \begin{small}
 \begin{equation}\label{deterqetaS1}
\det\left[\begin{smallmatrix}
z\mathbb{I}_{r}-A_0\big( \mathbb{I}_r-\overrightarrow{q}\overrightarrow{q}^* \big) & 
-(\sqrt{\mu}+\overline{y})\mathbb{I}_{r}
\\
(\sqrt{\mu}+y)\mathbb{I}_{r} & 
\overline{z}\mathbb{I}_{r}-A_0^*\big( \mathbb{I}_r-\overrightarrow{q}\overrightarrow{q}^* \big)
\end{smallmatrix}\right]
=
\det\left[\begin{smallmatrix}
z\mathbb{I}_{r}-J\big( \mathbb{I}_r
-\overrightarrow{\eta}\overrightarrow{\eta}^*U \big) & 
-(\sqrt{\mu}+\overline{y})U^{-1}
\\
(\sqrt{\mu}+y)U & 
\overline{z}\mathbb{I}_{r}-J^*\big( \mathbb{I}_r
-U\overrightarrow{\eta}\overrightarrow{\eta}^* \big)
\end{smallmatrix}\right],
 \end{equation}
 \end{small}
 and
   \begin{small}
 \begin{multline}\label{deterqetaS2}
\det\left[\begin{smallmatrix}
z\mathbb{I}_{r}-A_0\big( \mathbb{I}_r-\overrightarrow{q}\overrightarrow{q}^* \big) & 
-(\sqrt{\mu}+\overline{y})\mathbb{I}_{r}
\\
\sqrt{\mu} A_0^*\overrightarrow{q}\overrightarrow{q}^*A_0\big( \mathbb{I}_r-\overrightarrow{q}\overrightarrow{q}^* \big)+y\mathbb{I}_{r} & 
\overline{z}\mathbb{I}_{r}-A_0^*\big( \mathbb{I}_r-\overrightarrow{q}\overrightarrow{q}^* \big)
\end{smallmatrix}\right]
\\=
\det\left[\begin{smallmatrix}
z\mathbb{I}_{r}-J\big( \mathbb{I}_r
-\overrightarrow{\eta}\overrightarrow{\eta}^*U \big) & 
-(\sqrt{\mu}+\overline{y})U^{-1}
\\
\sqrt{\mu}J^*
U\overrightarrow{\eta}\overrightarrow{\eta}^*
UJ
 \big( \mathbb{I}_r
 -\overrightarrow{\eta}\overrightarrow{\eta}^*U \big)+yU & 
\overline{z}\mathbb{I}_{r}-J^*\big( \mathbb{I}_r
-U\overrightarrow{\eta}\overrightarrow{\eta}^* \big)
\end{smallmatrix}\right].
 \end{multline}
 \end{small}
 
To compute the terms $\partial /(\partial \mu)\det(\cdots)$ in \eqref{ImuI}, we can view the corresponding determinants as polynomials of $\mu^{\frac{1}{2}}$:
\begin{small}
\begin{equation*}
P\big(\mu^{\frac{1}{2}}\big)=a_0+a_1\mu^{\frac{1}{2}}
+a_2\mu+a_3\mu^{\frac{3}{2}}+\cdots.
\end{equation*}
\end{small}
Hence,
\begin{small}
\begin{equation}\label{partialPmu}
(\partial/\partial \mu)P\big( \mu^{\frac{1}{2}} \big)
=(a_1/2)\mu^{-\frac{1}{2}}+a_2+(3a_3/2)\mu^{\frac{1}{2}}+\cdots.
\end{equation}
\end{small}

From basic computations, we find the coefficient $a_1$ is of the form $(c_1 y+c_2\overline{y})\widetilde{P}(|y|^2)$, with $\widetilde{P}$ a polynomial, $c_1$, $c_2$ and the coefficients of $\widetilde{P}$ are independent from $y$ and $\overline{y}$. Note in \eqref{ImuI},
\begin{small}
\begin{equation*}
\int_{\mathbb{C}} (c_1 y+c_2\overline{y})\widetilde{P}(|y|^2)
e^{-\frac{N}{\tau}|y|^2}{\rm d}y=0,
\end{equation*}
\end{small}
so we can drop out the $(a_1/2)\mu^{-\frac{1}{2}}$ term in \eqref{partialPmu} before setting $\mu=0$ in \eqref{intermediatelemmaEqua} and \eqref{ImuI};

 Recalling \eqref{ImuI}, \eqref{g2 eta y}, \eqref{determinantSimplyfy1} and \eqref{deterqetaS1}, the coefficient $a_2$ equals to $g_2\big( \overrightarrow{\eta},y \big)$ when we are considering
 $\det\left[\begin{smallmatrix}
z\mathbb{I}_{N-1}-B_{2,2} & -(\sqrt{\mu}+\overline{y})\mathbb{I}_{N-1}
\\
(\sqrt{\mu}+y)\mathbb{I}_{N-1} & \overline{z}\mathbb{I}_{N-1}-B_{2,2}^*
\end{smallmatrix}\right]$; Recalling \eqref{ImuI}, \eqref{g3 eta y}, \eqref{determinantSimplyfy2} and \eqref{deterqetaS2}, the coefficient $a_2$ equals to $g_3\big( \overrightarrow{\eta},y \big)$ when we are considering
 $\det\left[\begin{smallmatrix}
z\mathbb{I}_{N-1}-B_{2,2} & -(\sqrt{\mu}+\overline{y})\mathbb{I}_{N-1}
\\
\sqrt{\mu}B_{1,2}B_{1,2}^*+y\mathbb{I}_{N-1} & \overline{z}\mathbb{I}_{N-1}-B_{2,2}^*
\end{smallmatrix}\right]$;

Finally, when we set $\mu=0$ in \eqref{intermediatelemmaEqua} and \eqref{ImuI}, the term $\sum_{k\geq 3}(ka_k/2)\mu^{\frac{k}{2}-1}$ in \eqref{partialPmu} vanishes.

Combining \eqref{intermediatelemmaEqua}-\eqref{Volume times Dirac},  \eqref{hatu1Integration} and the above arguments, we can complete the proof.

\end{proof}

\section{Proof of Theorem \ref{OverlapStaEdgeOutlier1}}\label{TheoremProofsOutlier}
In Proposition \ref{IntegralRepresent}, set $\rho=\frac{1}{2r}$ and $z=z_0+N^{-\frac{1}{2r}}\hat{z}$.

{\bf Step 1: Taylor expansion of $ f(\overrightarrow{\eta},y). $} In \eqref{IntegralRepresentEqua} and \eqref{f eta y}, set
\begin{equation*}
\overrightarrow{\eta}=\sqrt{s}\overrightarrow{\chi},\quad
\overrightarrow{\chi}^*P^*P\overrightarrow{\chi}=1,
\end{equation*}
with Jacobian
\begin{equation*}
{\rm d}\overrightarrow{\eta}=s^{r-1}\delta(1-\overrightarrow{\chi}^*P^*P\overrightarrow{\chi}){\rm d}s{\rm d}\overrightarrow{\chi}.
\end{equation*}

Further set $P=\begin{bmatrix}
P_{11} & P_{12} \\ P_{21} & P_{22}
\end{bmatrix}$, with $P_{11}$ a $r\times r$ matrix and $P_{22}$ a number. Further make the change of variable
\begin{equation*}
\begin{bmatrix}
\widehat{\chi}_2
\\
\vdots
\\
\widehat{\chi}_r
\end{bmatrix}=
\begin{bmatrix}
\chi_2
\\
\vdots
\\
\chi_r
\end{bmatrix}-N^{-\frac{1}{2r}}\hat{z}\begin{bmatrix}
\chi_1
\\
\vdots
\\
\chi_{r-1}
\end{bmatrix}
-N^{-\frac{1}{2r}}\hat{z}P_{11}^{-1}P_{12}\chi_{r},
\end{equation*}
and
\begin{equation*}
s=1-\tau/\big(
|z_0|^2+N^{-\frac{1}{2r}}(\overline{z}_0\hat{z}+z_0\overline{\hat{z}})
+N^{-\frac{1}{r}}|\hat{z}|^2
\big)+\widetilde{s}.
\end{equation*}
After basic computations we have
\begin{small}
\begin{multline*}
f(\overrightarrow{\eta},y)-\frac{1}{\tau}|z|^2=
\log(\tau)-1
-\frac{1}{\tau}
\big(
1-\frac{\tau}{|z_0|^2}
\big)N^{-1}
\big(
\vec{e}_1^*P^*P \vec{e}_1
\big)^{-1}
\big(
\vec{e}_r^*P^{-1}(P^*)^{-1} \vec{e}_r
\big)^{-1}
\big|\hat{z}\big|^{2r}        
-\frac{|z_0|^4}{2\tau^2}\widetilde{s}^2
\\
-\frac{1}{\tau}
\big(
1-\frac{\tau}{|z_0|^2}
\big)|y|^2-\frac{1}{\tau}
\big(
1-\frac{\tau}{|z_0|^2}
\big)\bigg(
\left[\begin{smallmatrix}
\widehat{\chi}_2
\\
\vdots
\\
\widehat{\chi}_r
\end{smallmatrix}\right]+
N^{-\frac{1}{2}}\hat{z}^r \widehat{\chi}_1
\frac{P_{21}P_{11}^{-1}P_{12}-P_{22}}{\vec{e}_1^*P^*P \vec{e}_1}
\big(
P_{11}^*P_{11}+P_{21}^*P_{21}
\big)^{-1}P_{21}^*
\bigg)^*
\\
\times
\big(
P_{11}^*P_{11}+P_{21}^*P_{21}
\big)\bigg(
\left[\begin{smallmatrix}
\widehat{\chi}_2
\\
\vdots
\\
\widehat{\chi}_r
\end{smallmatrix}\right]+
N^{-\frac{1}{2}}\hat{z}^r \widehat{\chi}_1
\frac{P_{21}P_{11}^{-1}P_{12}-P_{22}}{\vec{e}_1^*P^*P \vec{e}_1}
\big(
P_{11}^*P_{11}+P_{21}^*P_{21}
\big)^{-1}P_{21}^*
\bigg) 
\\+O\Big(
N^{-1}\big(|\hat{z}|+1 \big)^{2r}
\big(N^{-\frac{1}{2r}}\big(|\hat{z}|+1 \big)
+|\widetilde{s}|\big)
+N^{-\frac{r+1}{2r}}\big(|\hat{z}|+1 \big)^{r+1}
\| \widehat{\chi} \|_2
+\| \widehat{\chi} \|_2^2\big(N^{-\frac{1}{2r}}\big(|\hat{z}|+1 \big)
+|\widetilde{s}|\big)
\Big)
\\
+O\Big(
|y|^4+N^{-\frac{1}{2r}}\big(|\hat{z}|+1 \big)|y|^2
+N^{-1-\frac{1}{2r}}\big(|\hat{z}|+1 \big)
+|\widetilde{s}|^3+N^{-\frac{1}{2r}}\widetilde{s}^2\big(|\hat{z}|+1 \big)
\Big).
\end{multline*}
\end{small}
Here, $\| \widehat{\chi} \|_2:=\sum_{i=2}^r \| \widehat{\chi}_i \|$ and
\begin{small}
\begin{equation*}
\chi_1=\sqrt{1-\overline{\chi}_1 U_{12}\overrightarrow{\chi}_2-
\chi_1\overrightarrow{\chi}_2^*U_{12}^*-\overrightarrow{\chi}_2^*U_{22}\overrightarrow{\chi}_2}
u_{11}^{-\frac{1}{2}}\widehat{\chi}_1,\quad
P^*P=U:=\begin{bmatrix}
u_{11} & U_{12} \\
U_{12}^* & U_{22}
\end{bmatrix},\quad \overrightarrow{\chi}_2:=\begin{bmatrix}
\chi_2 \\
\vdots \\
\chi_r
\end{bmatrix},
\end{equation*}
\end{small}
with $U_{22}$ a $(r-1)\times (r-1)$ matrix, and $u_{11}$ a number.

{\bf Step 2: Taylor expansion of $ g(\overrightarrow{\eta},y). $}
In \eqref{g eta y}, we have
\begin{equation*}
1-\overrightarrow{\eta}^*U\overrightarrow{\eta}=
=
1-s\overrightarrow{\chi}^*P^*P\overrightarrow{\chi}
=1-s=\frac{\tau}{|z_0|^2}+O\big(
N^{-\frac{1}{2r}}\big(|\hat{z}|+1 \big)+\big|\widetilde{s}\big|
\big),
\end{equation*}
and 
\begin{equation*}
|z|^2+|y|^2=|z_0|^2+O\big(
N^{-\frac{1}{2r}}\big(|\hat{z}|+1 \big)
+|y|^2
\big).
\end{equation*}
Hence, we have
\begin{equation*}
\big(
1-\overrightarrow{\eta}^*U\overrightarrow{\eta}
\big)^{-r-1}
\big(|z|^2+|y|^2\big)^{-r-1}
=\tau^{-r-1}+O\big(
N^{-\frac{1}{2r}}\big(|\hat{z}|+1 \big)+\big|\widetilde{s}\big|
+|y|^2
\big).
\end{equation*}

In \eqref{g1 eta y}-\eqref{g3 eta y}, from elementary calculus and operations for matrices, we can get
\begin{equation*}
g_1(\overrightarrow{\eta},y)=O\Big(
N^{-\frac{1}{2r}}\big(|\hat{z}|+1 \big)+\big|\widetilde{s}\big|
+|y|+\| \widehat{\chi} \|_2
\Big),
\end{equation*}
\begin{equation*}
\frac{\tau}{N} g_2(\overrightarrow{\eta},y)=(1+N|y|^2)O\Big(
N^{-\frac{1}{2r}}\big(|\hat{z}|+1 \big)+\big|\widetilde{s}\big|
+|y|+\| \widehat{\chi} \|_2
\Big),
\end{equation*}
and
\begin{multline*}
g_3(\overrightarrow{\eta},y)=(1+N|z_0|^{-2}|y|^2)|z_0|^2
\big(\vec{e}_1^*P^*P \vec{e}_1 \big)\big(
\vec{e}_r^*P^{-1}(P^*)^{-1} \vec{e}_r
\big)\big(
1-\frac{\tau}{|z_0|^2}
\big)
\\+
(1+N|y|^2)O\Big(
N^{-\frac{1}{2r}}\big(|\hat{z}|+1 \big)+\big|\widetilde{s}\big|
+|y|+\| \widehat{\chi} \|_2
\Big).
\end{multline*}
Therefore, we have
\begin{multline*}
g(\overrightarrow{\eta},y)=\tau^{-r-1}(1+N|z_0|^{-2}|y|^2)|z_0|^2
\big(\vec{e}_1^*P^*P \vec{e}_1 \big)\big(
\vec{e}_r^*P^{-1}(P^*)^{-1} \vec{e}_r
\big)\big(
1-\frac{\tau}{|z_0|^2}
\big)
\\+
(1+N|y|^2)O\Big(
N^{-\frac{1}{2r}}\big(|\hat{z}|+1 \big)+\big|\widetilde{s}\big|
+|y|+\| \widehat{\chi} \|_2
\Big).
\end{multline*}

{\bf Step 3.} In \eqref{DN}, from Stirling's formula, we have 
\begin{equation*}
D_N e^{\frac{N}{\tau}|z|^2}=
\frac{N^{r+\frac{1}{2}}e^N \pi^{-r-2}}{\sqrt{2\pi}\tau^{N+1}}
\det(U)\big(
1+O(N^{-1})
\big).
\end{equation*}
Further make a rescaling of the integration variables
\begin{equation*}
\big(
\widetilde{s},y,\widehat{\chi}_2,\cdots,
\widehat{\chi}_r
\big)\rightarrow
N^{-\frac{1}{2}}\big(
\widetilde{s},y,\widehat{\chi}_2,\cdots,
\widehat{\chi}_r
\big).
\end{equation*}
By repeating the procedure of Laplace method, as having been done in the proof of Theorem \ref{OverlapStaEdge}, see Proposition \ref{foranalysis} and the derivations for equations \eqref{INDeltaDecom}-\eqref{INdeltaDecomF}, we can come to the conclusion.

\section{Proof of Theorem \ref{OverlapStaEdgeOutlier2}}\label{TheoremProofsOutlier2}
In Proposition \ref{IntegralRepresent}, set $\rho=\frac{1}{2}$ and $z=z_0+N^{-\frac{1}{2}}\hat{z}$.

{\bf Step 1: Taylor expansion of $ f(\overrightarrow{\eta},y). $} In \eqref{IntegralRepresentEqua} and \eqref{f eta y}, set
\begin{equation*}
\overrightarrow{\eta}=\sqrt{s}\overrightarrow{\chi},\quad
\overrightarrow{\chi}^*\overrightarrow{\chi}=1,
\end{equation*}
with Jacobian
\begin{equation*}
{\rm d}\overrightarrow{\eta}=s^{r-1}\delta(1-\overrightarrow{\chi}^*\overrightarrow{\chi}){\rm d}s{\rm d}\overrightarrow{\chi}.
\end{equation*}

Further make the change of variable
\begin{equation*}
s=1-\tau/\big(
|z_0|^2+N^{-\frac{1}{2r}}(\overline{z}_0\hat{z}+z_0\overline{\hat{z}}
\big)+\widetilde{s}.
\end{equation*}
After basic computations we have
\begin{equation*}
s^{r-1}=\big(
1-\frac{\tau}{|z_0|^2}
\big)^{r-1}+O\Big(
N^{-\frac{1}{2}}\big(|\hat{z}|+1 \big)
+|\widetilde{s}|
\Big),
\end{equation*}
and
\begin{small}
\begin{multline*}
f(\overrightarrow{\eta},y)-\frac{1}{\tau}|z|^2=
\log(\tau)-1
-\frac{1}{\tau}
\big(
1-\frac{\tau}{|z_0|^2}
\big)
N^{-1}
\big|\hat{z}\big|^{2}        
-\frac{|z_0|^4}{2\tau^2}\widetilde{s}^2
-\frac{1}{\tau}
\big(
1-\frac{\tau}{|z_0|^2}
\big)|y|^2 
\\
+O\Big(
|y|^4
+N^{-\frac{1}{2}}|y|^2
+N^{-\frac{3}{2}}\big(|\hat{z}|+1 \big)^3
+|\widetilde{s}|^3+N^{-\frac{1}{2}}\widetilde{s}^2\big(|\hat{z}|+1 \big)
\Big).
\end{multline*}
\end{small}

{\bf Step 2: Taylor expansion of $ g(\overrightarrow{\eta},y). $}
In \eqref{g eta y}, we have
\begin{equation*}
1-\overrightarrow{\eta}^*U\overrightarrow{\eta}=
=
1-s\overrightarrow{\chi}^*\overrightarrow{\chi}
=1-s=\frac{\tau}{|z_0|^2}+O\big(
N^{-\frac{1}{2}}\big(|\hat{z}|+1 \big)+\big|\widetilde{s}\big|
\big),
\end{equation*}
and 
\begin{equation*}
|z|^2+|y|^2=|z_0|^2+O\big(
N^{-\frac{1}{2}}\big(|\hat{z}|+1 \big)
+|y|^2
\big).
\end{equation*}
Hence, we have
\begin{equation*}
\big(
1-\overrightarrow{\eta}^*U\overrightarrow{\eta}
\big)^{-r-1}
\big(|z|^2+|y|^2\big)^{-r-1}
=\tau^{-r-1}+O\big(
N^{-\frac{1}{2}}\big(|\hat{z}|+1 \big)+\big|\widetilde{s}\big|
+|y|^2
\big).
\end{equation*}

In \eqref{g1 eta y}-\eqref{g3 eta y}, from elementary calculus and operations for matrices, we can get
\begin{multline*}
g_1(\overrightarrow{\eta},y)=
\big(
1-\frac{\tau}{|z_0|^2}
\big)^2|z_0|^2\big(
N^{-1}|\hat{z}|^2+|y|^2
\big)^{r-1}
\\+
O\Big(
\big(
N^{-1}(1+|\hat{z}|)^2+|y|^2
\big)^{r-1}\big(
N^{-\frac{1}{2}}(1+|\hat{z}|)+\big|\widetilde{s}\big|+|y|^2
\big)
\Big),
\end{multline*}
\begin{multline*}
\frac{\tau}{N} g_2(\overrightarrow{\eta},y)=
\tau \big(
1-\frac{\tau}{|z_0|^2}
\big)^2 \big(
1+N|z_0|^{-2}|y|^2
\big)\big(
N^{-1}|\hat{z}|^2+|y|^2
\big)^{r-1}
\\
+\tau \big(
1-\frac{\tau}{|z_0|^2}
\big)^2(r-1)\big(
N^{-1}|z_0|^2+2|y|^2
\big)\big(
N^{-1}|\hat{z}|^2+|y|^2
\big)^{r-2}
\\
+\tau (r-1)(r-2)|z_0|^{2}N^{-1}\big(
1-\frac{\tau}{|z_0|^2}
\big)^2|y|^2  
\big(
N^{-1}|\hat{z}|^2+|y|^2
\big)^{r-3}
\\
+\big(
1+N|y|^2
\big)O\Big(
\big(
N^{-1}(1+|\hat{z}|)^2+|y|^2
\big)^{r-1}\big(
N^{-\frac{1}{2}}(1+|\hat{z}|)+\big|\widetilde{s}\big|+|y|^2
\big)
\Big),
\end{multline*}
and
\begin{multline*}
g_3(\overrightarrow{\eta},y)=\tau \big(
1-\frac{\tau}{|z_0|^2}
\big)\big(
N^{-1}|\hat{z}|^2+|y|^2
\big)^{r-1}
\big(
1+N |z_0|^{-2}|y|^2
\big)
\\+
(r-1)\tau \big(
1-\frac{\tau}{|z_0|^2}
\big)|y|^2 \big(
N^{-1}|\hat{z}|^2+|y|^2
\big)^{r-2}
+O\Big(
\big(
N^{-1}(1+|\hat{z}|)^2+|y|^2
\big)^{r-1}\big(
N^{-\frac{1}{2}}(1+|\hat{z}|)+\big|\widetilde{s}\big|
\big)
\Big).
\end{multline*}
Therefore, we have
\begin{small}
\begin{multline*}
g(\overrightarrow{\eta},y)=\tau^{-r}
\big(
1-\frac{\tau}{|z_0|^2}
\big)\big(
N^{-1}|\hat{z}|^2+|y|^2
\big)^{r-1}
\Big(
\frac{|z_0|^2}{\tau}\big(
1-\frac{\tau}{|z_0|^2}
\big)+\big(
1+N |z_0|^{-2}|y|^2
\big)\big(
2-\frac{\tau}{|z_0|^2}
\big)
\\
+(r-1)\big(
1-\frac{\tau}{|z_0|^2}
\big)\big(
N^{-1} |z_0|^{2}+2|y|^2
\big)\big(
N^{-1}|\hat{z}|^2+|y|^2
\big)^{-1}+(r-1)(r-2)|z_0|^{2}N^{-1}\big(
1-\frac{\tau}{|z_0|^2}
\big)|y|^2  
\big(
N^{-1}|\hat{z}|^2+|y|^2
\big)^{-2}
\\+(r-1)|y|^2 \big(
N^{-1}|\hat{z}|^2+|y|^2
\big)^{-1}
\Big)+\big(
1+N|y|^2
\big)O\Big(
\big(
N^{-1}(1+|\hat{z}|)^2+|y|^2
\big)^{r-1}\big(
N^{-\frac{1}{2}}(1+|\hat{z}|)+\big|\widetilde{s}\big|+|y|^2
\big)
\Big).
\end{multline*}
\end{small}

{\bf Step 3.} In \eqref{DN}, from Stirling's formula, we have 
\begin{equation*}
D_N e^{\frac{N}{\tau}|z|^2}=
\frac{N^{r+\frac{1}{2}}e^N \pi^{-r-2}}{\sqrt{2\pi}\tau^{N+1}}
\big(
1+O(N^{-1})
\big).
\end{equation*}
Further make a rescaling of the integration variables
\begin{equation*}
\big(
\widetilde{s},y
\big)\rightarrow
N^{-\frac{1}{2}}\big(
\widetilde{s},y
\big).
\end{equation*}
By repeating the procedure of Laplace method, as having been done in the proof of Theorem \ref{OverlapStaEdge}, see Proposition \ref{foranalysis} and the derivations for equations \eqref{INDeltaDecom}-\eqref{INdeltaDecomF}, we can get
\begin{equation}\label{ONtauOutlier2Final02}
\mathcal{O}_N\big(
\tau,A_0,z_0+N^{-\frac{1}{2}}\hat{z}
\big)=\frac{\big(
1-\frac{\tau}{|z_0|^2}
\big)^r}{\pi\tau^r \Gamma(r)|z_0|^2}
e^{-\frac{1}{\tau}\big(
1-\frac{\tau}{|z_0|^2}
\big)|\hat{z}|^2}I_0
+O\big(
N^{-\frac{1}{2}}
(|\hat{z}|+1)^{2r+1}
\big),
\end{equation}
where 
\begin{multline*}
I_0=\int_0^{+\infty}
\big(
|\hat{z}|^2+u
\big)^{r-1}\Big(
\frac{|z_0|^2}{\tau}\big(
1-\frac{\tau}{|z_0|^2}
\big)+\big(
1+|z_0|^{-2}u
\big)\big(
2-\frac{\tau}{|z_0|^2}
\big)
\\
+(r-1)\big(
1-\frac{\tau}{|z_0|^2}
\big)\big(
|z_0|^{2}+2u
\big)\big(
|\hat{z}|^2+u
\big)^{-1}+(r-1)(r-2)|z_0|^{2}\big(
1-\frac{\tau}{|z_0|^2}
\big)u 
\big(
|\hat{z}|^2+u
\big)^{-2}
\\+(r-1)u \big(
|\hat{z}|^2+u
\big)^{-1}
\Big)
e^{-\frac{1}{\tau}\big(
1-\frac{\tau}{|z_0|^2}
\big)u}{\rm d}u.
\end{multline*}

For a real number $a$ and a positive integer $p$, set 
\begin{equation*}
J_p(a)=\int_0^{+\infty} (a+u)^{p-1}e^{-u}{\rm d}u,
\end{equation*}
from integration by parts, we have the following recurrence relation
\begin{equation*}
J_{p+1}(a)=pJ_p(a)+a^p, \quad p\geq 1.
\end{equation*}
Using this relation and basic computations we have
\begin{multline}\label{I0integralOutlier2}
I_0=\tau^{r-1}\big(
1-\frac{\tau}{|z_0|^2}
\big)^{-r-1}|z_0|^2\Gamma(r)
\\
\times\Big(
re_{r}\big(\frac{1}{\tau}\big(
1-\frac{\tau}{|z_0|^2}
\big)|\hat{z}|^2\big)-\frac{1}{\tau}\Big(
1-\frac{\tau}{|z_0|^2}
\big)|\hat{z}|^2e_{r-1}\big(\frac{1}{\tau}\big(
1-\frac{\tau}{|z_0|^2}
\big)|\hat{z}|^2\big)
\Big).
\end{multline}
Combining \eqref{ONtauOutlier2Final02} and \eqref{I0integralOutlier2}, we can come to the conclusion.

\section{Proofs of Propositions \ref{foranalysis}, \ref{fTaylor} and \ref{gTaylor}}\label{propproofs}

 \subsection{Proof of Proposition \ref{foranalysis}.}
Recalling \eqref{f0 eta y} and the maximum statement, Lemma \ref{maximumlemma}. We have for any $\delta>0$, there exists $\Delta>0$ such that
\begin{small}
\begin{equation}\label{foranalysis1}
f_0\big( \overrightarrow{\eta},y \big)-\log(\tau)
\leq -\Delta,\quad \textit{if}\ \ 
\overrightarrow{\eta}^*U\overrightarrow{\eta}+|y|^2>\delta
\ \  \textit{and}\ \ \overrightarrow{\eta}^*\overrightarrow{\eta}\leq 1.
\end{equation}
\end{small}
As in \eqref{IntegralRepresentEqua}, we have $\overrightarrow{\eta}^*\overrightarrow{\eta}\leq 1,$ also $\tau>0$. Hence, in \eqref{f eta y} and \eqref{logzyrewrite}, we have
\begin{small}
\begin{equation}\label{foranalysis2}
\frac{1}{\tau}
N^{-\frac{1}{2}}\big( \hat{z}\overrightarrow{\eta}^*J^*U\overrightarrow{\eta}
+\overline{\hat{z}}\overrightarrow{\eta}^*UJ\overrightarrow{\eta}
 \big)=O\big(
 N^{-\frac{1}{2}}
 \big),
\end{equation}
\end{small}
 and
\begin{small}
\begin{equation}\label{foranalysis3}
\log\big(
1+(\tau+|y|^2)^{-1}\big(
(z_0\overline{\hat{z}}+\overline{z}_0\hat{z})N^{-\frac{1}{2}}
+N^{-1}|\hat{z}|^2
\big)
\big)=O\big(
 N^{-\frac{1}{2}}
 \big).
\end{equation}
\end{small}
 
For $g\big( \overrightarrow{\eta},y \big)$, as $\tau>0$, when $N$ is sufficiently large, \begin{small}
$\big( |z|^2+|y|^2 \big)^{-r-3}=O(1)$,
\end{small} the remaining factor is a polynomial of finite degree with coefficients bounded by $O(N)$. So by taking a sufficiently large $N_0>r+3$, we have
\begin{small}
\begin{equation}\label{foranalysis4}
\int_{\Omega_{\delta}^{c}\cap
 \{ \overrightarrow{\eta}^*U\overrightarrow{\eta}\leq 1 \}}
| g\big( \overrightarrow{\eta},y \big) |\exp\Big\{ N_0\Big(
 f_0\big( \overrightarrow{\eta},y \big)-\log(\tau) \Big) \Big\}
{\rm d}\overrightarrow{\eta}{\rm d}y
=O(N).
\end{equation}
\end{small}
Hence, combining \eqref{foranalysis1}-\eqref{foranalysis4}, we have
\begin{small}
\begin{multline}
\bigg|
\int_{\Omega_{\delta}^{c}\cap
 \{ \overrightarrow{\eta}^*U\overrightarrow{\eta}\leq 1 \}}
g\big( \overrightarrow{\eta},y \big)\exp\Big\{ 
N\Big( f\big( \overrightarrow{\eta},y \big)-\log(\tau) \Big) \Big\}
{\rm d}\overrightarrow{\eta}{\rm d}y
\bigg|   \\
\leq e^{-(N-N_0)\Delta+O(\sqrt{N})}
 O(N)    =O\big(
e^{-\frac{1}{2}N\Delta}
\big).
\end{multline}
\end{small}

This  thus completes the proof.

 \subsection{Proof of Proposition \ref{fTaylor}.}
In \eqref{f eta y}, we have assumed that $|z_0|^2=\tau$ and $\rho=\frac{1}{2}$. First we have
\begin{small}
\begin{equation}\label{flogexpan}
\log\big(
1-\overrightarrow{\eta}^*U\overrightarrow{\eta}
\big)+\overrightarrow{\eta}^*U\overrightarrow{\eta}
=-\big( \overrightarrow{\eta}^*U\overrightarrow{\eta} \big)^2/2
+O\big( \| \overrightarrow{\eta} \|^6 \big).
\end{equation}
\end{small}
Setting
\begin{small}
\begin{equation}\label{widetildeU}
\widetilde{U}=\big( z_0\mathbb{I}_r-J \big)^*U\big( z_0\mathbb{I}_r-J \big),
\end{equation}
\end{small}
 According to   the Jordan block structure of  $J$ in \eqref{Jdetail}, both $U$ and $\widetilde{U}$ admit  a three-layer  structure
 \begin{small}
 \begin{equation}\label{Ublockform1}
U=[ U_{k;l} ]_{k,l=1}^{m+1},  
\quad \widetilde{U}=[ \widetilde{U}_{k;l} ]_{k,l=1}^{m+1},
\end{equation}
 \end{small}
where $U_{m+1;m+1}$ and $\widetilde{U}_{m+1;m+1}$ are of sizes $r_0\times r_0$ with $r_0$ defined in \eqref{geometricmultiplicity}. \eqref{Ublockform1} corresponds to the outer direct sum \begin{small}
$
(\bigoplus_{i=1}^m (...))
\bigoplus J_1$
\end{small} in \eqref{Jdetail}, and for $k,l=1,\cdots,m$,
\begin{small}
\begin{equation}\label{Ublockform2}
\begin{aligned}
&U_{k;l}=\left[ U_{k,a;l,b} \right]_{n_{k}\times n_{l}},\quad
\widetilde{U}_{k;l}=\left[ \widetilde{U}_{k,a;l,b} \right], \\
&
U_{m+1;l}=\left[ U_{m+1;l,b} \right]_{1\times  n_{l}}, 
 \quad
\widetilde{U}_{m+1;l}=\left[ \widetilde{U}_{m+1;l,b}  \right]_{1\times  n_{l}}, \\
&U_{k;m+1}=\left[ U_{k,a;m+1} \right]_{n_{k} \times  1},\quad
\widetilde{U}_{k;m+1}=\left[ \widetilde{U}_{k,a;m+1} \right]_{n_{k} \times  1},
\end{aligned}
\end{equation}
\end{small}
 with   $U_{k,a;l,b}$, $U_{m+1;l,b}$ and $U_{k,a;m+1}$ being  of sizes $p_{k}\times p_{l}$, $r_0\times p_{l}$ and $p_{k}\times r_0$ respectively, and the same for 
 $\widetilde{U}_{k,a;l,b}$, $\widetilde{U}_{m+1;l,b}$ and $\widetilde{U}_{k,a;m+1}$. The 1st, 3rd and 5th equations in \eqref{Ublockform2} corresponds to the inner direct sum  sum \begin{small}
$R_{p_{i}}(z_0 )   \bigoplus \cdots \bigoplus  R_{p_{i}}(z_0 )$ 
\end{small} in \eqref{Jdetail}, and the same for the 2nd, 4th and 6th equations.
 
Recalling \eqref{Jdetail}, we know that after minusing $J_1$, the diagonal of $J$ is all $z_0$, so from the relation \eqref{widetildeU} we have
 \begin{small}
\begin{multline}\label{UtildeUrelation1}
\widetilde{U}_{k,a;l,b}=R_{p_{k}}(0)^*U_{k,a;l,b}R_{p_{l}}(0), 
\quad
\widetilde{U}_{m+1;l,b}=-\big( z_0\mathbb{I}_{r_0}-J_1 \big)^*U_{m+1;l,b}
R_{p_{l}}(0),  \\ 
\widetilde{U}_{k,a;m+1}=-R_{p_{k}}(0)^*U_{k,a;m+1}\big( z_0\mathbb{I}_{r_0}-J_1 \big),
\quad
\widetilde{U}_{m+1;m+1}=\big( z_0\mathbb{I}_{r_0}-J_1 \big)^*
U_{m+1;m+1}\big( z_0\mathbb{I}_{r_0}-J_1 \big);              
\end{multline}
\end{small}
 In order to be able to utilize the structure of  $R_{p_{k}}(0)$, we need corresponding block structures for the quantities $U_{k,a;l,b}$, $U_{m+1;l,b}$ and $U_{k,a;m+1}$ in \eqref{Ublockform2}, to this end, for $k,l=1,\cdots,m$ set 
\begin{small}
\begin{equation}\label{Ublockform3}
U_{k,a;l,b}=\begin{bmatrix}
U_{k,a;l,b}^{(11)}& U_{k,a;l,b}^{(12)} \\
U_{k,a;l,b} ^{(21)}& U_{k,a;l,b}^{(22)}
\end{bmatrix},        \quad
U_{k,a;m+1}=\begin{bmatrix}
U_{k,a;m+1}^{(U)} \\ U_{k,a;m+1}^{(D)}
\end{bmatrix},
U_{m+1;l,b}=\begin{bmatrix}
U_{m+1;l,b}^{(L)} & U_{m+1;l,b}^{(R)}
\end{bmatrix},
\end{equation}
\end{small}
where $U_{k,a;l,b}^{(11)}$, $U_{m+1;l,b}^{(L)}$ and $U_{k,a;m+1}^{(U)}$ are of sizes $(p_{k}-1)\times(p_{l}-1)$, ${r_0}\times(p_{l}-1)$ and $(p_{k}-1)\times {r_0}$ respectively.
From \eqref{1Rpirealedge} we know that the matrix $R_{p_{k}}(0)$ is actually the shift operator, simple calculations show that 
\begin{small}
\begin{multline}\label{UtildeUrelation2}
\widetilde{U}_{k,a;l,b}=\begin{bmatrix}
0 & 0 \\
0 & U_{k,a;l,b}^{(11)}
\end{bmatrix},   \quad  \widetilde{U}_{k,a;m+1}^{(1)}=\begin{bmatrix}
0 \\ -U_{k,a;m+1}^{(U)}\big( z_0\mathbb{I}_{r_0}-J_1 \big)
\end{bmatrix},
  \\
\widetilde{U}_{m+1;l,b}=\begin{bmatrix}
0 & -\big( z_0\mathbb{I}_{r_0}-J_1 \big)^*D_{m+1;l,b}^{(L)}
\end{bmatrix},\quad
\widetilde{U}_{m+1,m+1}=U_{m+1,m+1}.
\end{multline}
\end{small}

Recalling \eqref{widetildeU}, in \eqref{f eta y} we have
\begin{small}
\begin{equation}\label{etaUeta}
\overrightarrow{\eta}^*
\big( z_0\mathbb{I}_r-J \big)^*U\big( z_0\mathbb{I}_r-J \big)\overrightarrow{\eta}
=\overrightarrow{\eta}^*
\widetilde{U}\overrightarrow{\eta}.
\end{equation}
\end{small}
 To compute this quadratic form explicitly, for the vector variable 
 $\overrightarrow{\eta}$, we need a block structure corresponding to $\widetilde{U}$ in \eqref{UtildeUrelation2}. To this end, put
\begin{equation*}\label{etablock1}
\overrightarrow{\eta}=\begin{bmatrix}
\overrightarrow{\eta_1} & \cdots & \overrightarrow{\eta_m} & 
\overrightarrow{\eta}_{m+1}
\end{bmatrix},
\end{equation*}
where $\overrightarrow{\eta}_{m+1}$ is of size $1\times r_0$ and for $l=1,\cdots,m$ 
\begin{small}
\begin{equation*}\label{etablock2}
\overrightarrow{\eta_l}=[
\overrightarrow{\eta_{l,b}}]_{1\times  n_{l}},\quad
\overrightarrow{\eta_{l,b}}=\begin{bmatrix}
\overrightarrow{\eta_{l,b}}^{(L)} & \overrightarrow{\eta_{l,b}}^{(R)}
\end{bmatrix},
\end{equation*}
\end{small}
while $\overrightarrow{\eta_{l,b}}$ and $\overrightarrow{\eta_{l,b}}^{(L)} $ are of sizes $1\times p_{l}$ and $1\times 1$ respectively. Now we can draw some blocks from 
$\overrightarrow{\eta}$ and  restructure it into  two new vectors
\begin{small}
\begin{equation}\label{etaUcompression1}
\overrightarrow{\eta_R}=\begin{bmatrix}
\overrightarrow{\eta_{1}}^{(R)} & \cdots & 
\overrightarrow{\eta_{m}}^{(R)} & \overrightarrow{\eta}_{m+1}
\end{bmatrix}, \ 
\overrightarrow{\eta_L}=\begin{bmatrix}
\overrightarrow{\eta_{1}}^{(L)} & \cdots & 
\overrightarrow{\eta_{m}}^{(L)}
\end{bmatrix},\quad
\overrightarrow{\eta_{l}}^{(L)}=[\overrightarrow{\eta_{l,b}}^{(L)}],
\ 
\overrightarrow{\eta_{l}}^{(R)}=[\overrightarrow{\eta_{l,b}}^{(R)}],
\end{equation}
\end{small}
Note that the column block vectors $\overrightarrow{\eta_R}$ and $\overrightarrow{\eta_L}$ are in fact a partition of the vector $\overrightarrow{\eta}$. The reason for the construction of \eqref{etaUcompression1} comes from that in \eqref{etaUeta} the quadratic form with respect to the matrix variable is degenerated, as there are columns of zeros in $\widetilde{U}$, see \eqref{UtildeUrelation2}. $\overrightarrow{\eta_R}$ corresponds exactly to the non-singular part of $\widetilde{U}$, while $\overrightarrow{\eta_L}$ corresponds exactly to the zeros in $\widetilde{U}$. Hence, $\overrightarrow{\eta_L}$ disappears in the quadratic form, and the $\overrightarrow{\eta_R}$ are all left.

With the above arguments and \eqref{UtildeUrelation1} in mind, define 
\begin{small}
\begin{equation} \label{Ut22R}
 \widetilde{U}^{(22)}=
 \big[\widetilde{U}^{(22)}_{k,l}\big]_{k,l=1}^{m+1},
\end{equation} 
\end{small}
    where for $k,l=1,\cdots,m$,
    \begin{small}
\begin{multline}\label{QUcompressiondetail1}
\widetilde{U}_{k,l}^{(22)}=\big[
U_{k,a;l,b}^{(22)}
\big], \quad
\widetilde{U}_{m+1,l}^{(22)}=\begin{bmatrix}
-\left( z_0\mathbb{I}_{r_0}-J_1 \right)^*U_{m+1;l,b}^{(L)}
\end{bmatrix}, 
\\
\widetilde{U}_{k,m+1}^{(22)}=\big[
-U_{k,a;m+1}^{(U)}\left( z_0\mathbb{I}_{r_0}-J_1 \right)
\big],\quad 
\widetilde{U}_{m+1,m+1}^{(22)}=\widetilde{U}_{m+1,m+1}.
\end{multline}
\end{small}
Compared to \eqref{UtildeUrelation1}, we discover that $\widetilde{U}^{(22)}$ is in fact a compression of the matrix $\widetilde{U}$ in \eqref{widetildeU} by omitting all zeros in \eqref{UtildeUrelation2}. After extracting out the invertible matrices 
$-\big( z_0\mathbb{I}_{r_0}-J_1 \big)^*$ and etc. from both sides in \eqref{QUcompressiondetail1}, the remaining matrix in $\widetilde{U}^{(22)}$ is in fact a principle sub-matrix of $U$, which is strictly positive from \eqref{U}, so $\widetilde{U}^{(22)}$ is strictly positive.

Now, combining \eqref{etaUeta} and the arguments before \eqref{Ut22R}, we have
\begin{small}
\begin{equation}\label{Tretatildeta}
\overrightarrow{\eta}^*
\big( z_0\mathbb{I}_r-J \big)^*U\big( z_0\mathbb{I}_r-J \big)\overrightarrow{\eta}
=\overrightarrow{\eta_R}^*
\widetilde{U}^{(22)}\overrightarrow{\eta_R}.
\end{equation}
\end{small}

Now, if we rewrite $J$ as
\begin{equation}\label{JJJJJJ}
J:=z_0\mathbb{I}_{r-r_0}\bigoplus 0_{r_0}+\widetilde{J},
\end{equation}
that is, in view of \eqref{Jdetail}, for $i=1,\cdots,m$, extract the diagonal part and denote the remaining part of $J$ as $\widetilde{J}$. Then from \eqref{etaUcompression1} we arrive at 
\begin{small}
\begin{equation}\label{J2jerrorcontrol1}
\overrightarrow{\eta}^*U\widetilde{J}
\overrightarrow{\eta}=O\big(
\big\| \overrightarrow{\eta_R} \big\|
\big\| \overrightarrow{\eta_L} \big\|
\big),\quad
\overrightarrow{\eta}^*U\big(J-\widetilde{J}\big)
\overrightarrow{\eta}=
z_0+\overrightarrow{\eta_L}^*
U[I_1]
\overrightarrow{\eta_L}
O\big(
\big\| \overrightarrow{\eta_R} \big\|
\big\| \overrightarrow{\eta_L} \big\|
\big),
\end{equation}
\end{small}
from which 
\begin{small}
\begin{equation}\label{J2jerrorcontrol2}
N^{-\frac{1}{2}}\big( \hat{z}\overrightarrow{\eta}^*J^*U\overrightarrow{\eta}
+\overline{\hat{z}}\overrightarrow{\eta}^*UJ\overrightarrow{\eta}
 \big)/\tau
 =N^{-\frac{1}{2}}\big( \overline{z}_0\hat{z}
+z_0\overline{\hat{z}}
 \big)\overrightarrow{\eta_L}^*
U[I_1]
\overrightarrow{\eta_L}/\tau.
\end{equation}
\end{small}
In \eqref{flogexpan}, in the same spirit we also have
\begin{small}
\begin{equation}\label{etaUetaetaL}
\overrightarrow{\eta}^*U\overrightarrow{\eta}
=\overrightarrow{\eta_L}^*
U[I_1]
\overrightarrow{\eta_L}+O\big(
\big\| \overrightarrow{\eta_R} \big\|
\big\| \overrightarrow{\eta_L} \big\|
\big),
\end{equation}
\end{small}
so
\begin{small}
\begin{equation}\label{logetaUetaetaL}
\log\big(
1-\overrightarrow{\eta}^*U\overrightarrow{\eta}
\big)+\overrightarrow{\eta}^*U\overrightarrow{\eta}
=-\big( 
\overrightarrow{\eta_L}^*
U[I_1]
\overrightarrow{\eta_L}
 \big)^2/2
+O\big( \| \overrightarrow{\eta} \|^6
+\| \overrightarrow{\eta} \|^3\| \overrightarrow{\eta_R} \| \big).
\end{equation}
\end{small}

Also we have
\begin{small}
\begin{multline}\label{yExpan}
-|y|^2/\tau+\log\big( |z|^2+|y|^2 \big)
=\log(\tau)+N^{-\frac{1}{2}}\big( \overline{z}_0\hat{z}
+z_0\overline{\hat{z}}
 \big)/\tau
 \\
+N^{-1}|\hat{z}|^2/\tau-
\big(
|y|^2+N^{-\frac{1}{2}}\big( \overline{z}_0\hat{z}
+z_0\overline{\hat{z}}
 \big)
\big)^2/(2\tau^2)+
O\big(
N\big(
|y|^2+N^{-\frac{1}{2}}
\big)^3
\big).
\end{multline}
\end{small}

Now combining \eqref{f eta y}, \eqref{Tretatildeta}, \eqref{J2jerrorcontrol2}, \eqref{logetaUetaetaL} and \eqref{yExpan}, we can come to the conclusion.

 \subsection{Proof of Proposition \ref{gTaylor}.}
For the Taylor expansion for $g\big( \overrightarrow{\eta},y \big)$ in \eqref{g eta y}, for the factor outside the round brackets, from \eqref{EdgeSetting} we have
\begin{small}
\begin{equation}\label{FirstFactor0}
\big(
1-\overrightarrow{\eta}^*U\overrightarrow{\eta}
\big)^{-r-1}=1+O\big(
\| \overrightarrow{\eta} \|^2
\big)
\end{equation}
\end{small}
and
\begin{small}
\begin{equation}\label{FirstFactor}
\big(|z|^2+|y|^2\big)^{-1}=\tau^{-1}+
O\big( N^{-\frac{1}{2}}+|y|^2 \big),\quad
\big(|z|^2+|y|^2\big)^{-r-1}
=\tau^{-r-1}+
O\big( N^{-\frac{1}{2}}+|y|^2 \big).
\end{equation}
\end{small}

To deal with the determinants in $g_i(\overrightarrow{\eta},y)$ for 
$i=1,2,3$(cf. \eqref{g1 eta y}-\eqref{g3 eta y}), from \eqref{Jdetail} and  \eqref{EdgeSetting} we can rewrite
\begin{small}
\begin{multline}\label{deterJ1}
z\mathbb{I}_r-J\big(
\mathbb{I}_r-\overrightarrow{\eta}\overrightarrow{\eta}^*U
\big)=z_0\mathbb{I}_r-J+O\big( N^{-\frac{1}{2}}
+\| \overrightarrow{\eta} \|^2 \big),
\\
z_0\mathbb{I}_r-J=\Big(-\bigoplus_{i=1}^m 
\overbrace{ R_{p_{i}}\left(0 \right)   \bigoplus \cdots \bigoplus  
R_{p_{i}}\left(0\right)
}^{n_{i}\ \mathrm{blocks}}\Big)\bigoplus \big(z_0\mathbb{I}_{r_0}-J_1\big),
\end{multline}
\end{small}
Also,
\begin{small}
\begin{multline}\label{deterJ2}
\overline{z}\mathbb{I}_r-J^*\big(
\mathbb{I}_r-U\overrightarrow{\eta}\overrightarrow{\eta}^*
\big)=
\Big(-\bigoplus_{i=1}^m 
\overbrace{ R_{p_{i}}\left(0 \right)^*   \bigoplus \cdots \bigoplus  
R_{p_{i}}\left(0\right)^*
}^{n_{i}\ \mathrm{blocks}}\Big)\bigoplus \big(\overline{z}_0\mathbb{I}_{r_0}-J_1^*\big)+
O\big( N^{-\frac{1}{2}}
+\| \overrightarrow{\eta} \|^2 \big).
\end{multline}
\end{small}
Applying Laplace expansion to cancel out the non-singular blocks: $z_0\mathbb{I}_{r_0}-J_1$, $\overline{z}_0\mathbb{I}_{r_0}-J_1^*$ and the $\mathbb{I}_{p_i-1}$ in $R_{p_{i}}\left(0 \right)$ and $R_{p_{i}}\left(0 \right)^*$, we have
\begin{small}
\begin{multline}\label{deterJ3}
\det\left[\begin{smallmatrix}
z\mathbb{I}_r-J\big( \mathbb{I}_r-\overrightarrow{\eta}\overrightarrow{\eta}^*U \big)       &
-\overline{y}U^{-1}  \\
yU  &  
\overline{z}\mathbb{I}_r-J^*\big( \mathbb{I}_r-U\overrightarrow{\eta}\overrightarrow{\eta}^* \big)
\end{smallmatrix}\right]=
\\
\Big(
\big| \det(z_0\mathbb{I}_{r_0}-J_1) \big|^2+
O\big( N^{-\frac{1}{2}}
+\| \overrightarrow{\eta} \|^2 \big)
\Big)
\det\Big(
\left[\begin{smallmatrix}
 &
-\overline{y}U^{-1}[I_2]  \\
yU[I_1]  &  
\end{smallmatrix}\right]
+
O\big( N^{-\frac{1}{2}}
+\| \overrightarrow{\eta} \|^2 \big)
\Big)
\\
=|y|^{2t}\det\big(
U[I_1]
\big) \det\big(
U^{-1}[I_2]
\big) \big| \det(z_0\mathbb{I}_{r_0}-J_1) \big|^2
+O\Big(
\big(
|y|+N^{-\frac{1}{2}}+\| \overrightarrow{\eta} \|^2
\big)^{2t}-|y|^{2t}
\Big).
\end{multline}
\end{small}
For the determinant 
\begin{small}
\begin{equation}\label{deterJ4}
\det\left[\begin{smallmatrix}
\big(z\mathbb{I}_r-J\big( \mathbb{I}_r-\overrightarrow{\eta}\overrightarrow{\eta}^*U \big)\big)
\big[ [r];[r]/\{ \beta \} \big]       &
-\overline{y}U^{-1}  \\
yU\big[ [r]/\{ \alpha \};[r]/\{ \beta \} \big]  &  
\big(\overline{z}\mathbb{I}_r-J^*\big( \mathbb{I}_r-U\overrightarrow{\eta}\overrightarrow{\eta}^* \big)\big)
\big[ [r]/\{ \alpha \};[r] \big]
\end{smallmatrix}\right]
\end{equation}
\end{small}
If $\alpha,\beta\in I_1$, there exists $0\leq k_1,k_2 <m$, $0\leq l_1<n_{k_1+1}$ and $0\leq l_2<n_{k_2+1}$, such that
\begin{small}
\begin{equation}\label{indexResentation}
\alpha=\sum_{i=1}^{k_1}p_i n_i+p_{k_1+1}l_1+1,
\quad
\beta=\sum_{i=1}^{k_2}p_i n_i+p_{k_2+1}l_2+1.
\end{equation}
\end{small}
Also introduce
\begin{small}
\begin{equation}\label{indexResentationTilde}
\widetilde{\alpha}=\sum_{i=1}^{k_1} n_i+l_1+1,
\quad
\widetilde{\beta}=\sum_{i=1}^{k_2} n_i+l_2+1,
\end{equation}
\end{small}
which corresponds to the position of $\alpha$ and $\beta$ in $I_1$ with increasing order.

After applying Laplace expansion to cancel out the non-singular part, we get that the determinant in \eqref{deterJ4} equals to
\begin{small}
\begin{multline}\label{deterJ5}
(-1)^{N(\alpha,\beta)}
\Big(
\big| \det(z_0\mathbb{I}_{r_0}-J_1) \big|^2+
O\big( N^{-\frac{1}{2}}
+\| \overrightarrow{\eta} \|^2 \big)
\Big)
\\
\times
\det\Big(
\left[\begin{smallmatrix}
 &
-\overline{y}U^{-1}[I_2]  \\
yU[I_1/\{ \widetilde{\alpha}\};I_1/\{ \widetilde{\beta} \}]  &  
\end{smallmatrix}\right]
+
O\big( N^{-\frac{1}{2}}
+\| \overrightarrow{\eta} \|^2 \big)
\Big),
\end{multline}
\end{small}
here
\begin{small}
\begin{equation}\label{IndexN}
N(\alpha,\beta)=(n_{k_1+1}-l_1)(p_{k_1+1}-1)+\sum_{i=k_1+2}^m n_i(p_i-1)
+r_0+\sum_{i=1}^{k_2} n_i(p_i-1)+l_2(p_{k_2+1}-1).
\end{equation}
\end{small}
Applying \eqref{geometricmultiplicity}, \eqref{indexResentation} and \eqref{indexResentationTilde}, we have
\begin{small}
\begin{equation}\label{IndexEquivalence}
(-1)^{\alpha+\beta+r+N(\alpha,\beta)}
=(-1)^{\widetilde{\alpha}+\widetilde{\beta}+t}.
\end{equation}
\end{small}

Now combining \eqref{deterJ5} and \eqref{IndexEquivalence}, we have if $\alpha,\beta\in I_1$,
\begin{small}
\begin{multline}\label{deterJ6}
(-1)^{\alpha+\beta+r}
\det\left[\begin{smallmatrix}
\big(z\mathbb{I}_r-J\big( \mathbb{I}_r-\overrightarrow{\eta}\overrightarrow{\eta}^*U \big)\big)
\big[ [r];[r]/\{ \beta \} \big]       &
-\overline{y}U^{-1}  \\
yU\big[ [r]/\{ \alpha \};[r]/\{ \beta \} \big]  &  
\big(\overline{z}\mathbb{I}_r-J^*\big( \mathbb{I}_r-U\overrightarrow{\eta}\overrightarrow{\eta}^* \big)\big)
\big[ [r]/\{ \alpha \};[r] \big]
\end{smallmatrix}\right]
\\
=(-1)^{\widetilde{\alpha}+\widetilde{\beta}}
|y|^{2(t-1)}\overline{y}
\big| \det(z_0\mathbb{I}_{r_0}-J_1) \big|^2
\det\big(
U[I_1/\{ \widetilde{\alpha}\};I_1/\{ \widetilde{\beta} \}]
\big)\det\big(
U^{-1}[I_2] 
\big)
\\
+O\Big(
\big(
|y|+N^{-\frac{1}{2}}+\| \overrightarrow{\eta} \|^2
\big)^{2t-1}-|y|^{2t-1}
\Big).
\end{multline}
\end{small}

If $\alpha\not\in I_1$ or $\beta\not\in I_1$, then the rank of non-singular parts in \eqref{deterJ4} will decrease, and such determinant will be contained in \begin{small}
$O\Big(
\big(
|y|+N^{-\frac{1}{2}}+\| \overrightarrow{\eta} \|^2
\big)^{2t}
\Big)$.
\end{small}

The same arguments for the determinant \eqref{deterJ4} can be applied to 
\begin{small}
\begin{equation*}
\det\left[\begin{smallmatrix}
\big(z\mathbb{I}_r-J\big( \mathbb{I}_r-\overrightarrow{\eta}\overrightarrow{\eta}^*U \big)\big)
\big[ [r]/\{ \alpha \};[r] \big]       &
-\overline{y}U^{-1}\big[ [r]/\{ \alpha \};[r]/\{ \beta \} \big]  \\
yU  &  
\big(\overline{z}\mathbb{I}_r-J^*\big( \mathbb{I}_r-U\overrightarrow{\eta}\overrightarrow{\eta}^* \big)\big)
\big[ [r];[r]/\{ \beta \} \big]
\end{smallmatrix}\right].
\end{equation*}
\end{small}
Also applying the same arguments we have that
\begin{small}
\begin{multline}\label{deterJ60}
\det\left[\begin{smallmatrix}
\big(z\mathbb{I}_r-J\big( \mathbb{I}_r-\overrightarrow{\eta}\overrightarrow{\eta}^*U \big)\big)
\big[ [r]/\{ \alpha_2 \};[r]\{ \beta_1 \} \big]       &
-\overline{y}U^{-1}\big[ [r]/\{ \alpha_2 \};[r]/\{ \beta_2 \} \big]  \\
yU\big[ [r]/\{ \alpha_1 \};[r]\{ \beta_1 \} \big]  &  
\big(\overline{z}\mathbb{I}_r-J^*\big( \mathbb{I}_r-U\overrightarrow{\eta}\overrightarrow{\eta}^* \big)\big)
\big[ [r]/\{ \alpha_1 \};[r]/\{ \beta_2 \} \big]
\end{smallmatrix}\right]
\\=
O\Big(
\big(
|y|+N^{-\frac{1}{2}}+\| \overrightarrow{\eta} \|^2
\big)^{2t-2}
\Big).
\end{multline}
\end{small}

Now in \eqref{g1 eta y} and \eqref{g2 eta y}, combining \eqref{FirstFactor}, \eqref{deterJ1}, \eqref{deterJ6} and \eqref{deterJ60}, we get that 
\begin{small}
\begin{equation}\label{g1 eta y Analysis}
g_1(\overrightarrow{\eta},y)
=O\Big(
\big(
|y|+N^{-\frac{1}{2}}+\| \overrightarrow{\eta} \|^2
\big)^{2t}
\Big)
\end{equation}
\end{small}
and
\begin{small}
\begin{multline}\label{g2 eta y Analysis}
(\tau/N)g_2(\overrightarrow{\eta},y)
=(N/\tau)
\big| \det(z_0\mathbb{I}_{r_0}-J_1) \big|^2
|y|^{2(t+1)}\det\big(
U[I_1]
\big)\det\big(
U^{-1}[I_2] 
\big)
\\+O\Big(
\big(
N|y|^2+1
\big)
\Big(
\big(
|y|+N^{-\frac{1}{2}}+\| \overrightarrow{\eta} \|^2
\big)^{2t}-|y|^{2t}
\Big)+|y|^{2t}+N|y|^{2(t+2)}
\Big).
\end{multline}
\end{small}

For $g_3(\overrightarrow{\eta},y)$ defined in \eqref{g3 eta y}, firstly we have a rough estimation
\begin{small}
\begin{equation}\label{rough estimation det}
J^*U\overrightarrow{\eta}
\overrightarrow{\eta}^*UJ
\big(
\mathbb{I}_r-\overrightarrow{\eta}
\overrightarrow{\eta}^*U
\big)=O\big(
\| \overrightarrow{\eta} \|^2
\big).
\end{equation}
\end{small}
For $\alpha,\beta\in I_1,$ according to the structure of $J$ and $\overrightarrow{\eta}$(cf. \eqref{Jdetail} and \eqref{etaUcompression1}), we have
\begin{small}
\begin{equation}\label{g3Est1}
\Big(
J^*U\overrightarrow{\eta}
\overrightarrow{\eta}^*UJ
\big(
\mathbb{I}_r-\overrightarrow{\eta}
\overrightarrow{\eta}^*U
\big)
\Big)_{\alpha,\beta}
=|z_0|^2\big(
U[I_1]\overrightarrow{\eta_L}
\overrightarrow{\eta_L}^*U[I_1]
\big)_{\widetilde{\alpha},\widetilde{\beta}}
+O\big(
\| \overrightarrow{\eta} \|\| \overrightarrow{\eta_R} \|
+\| \overrightarrow{\eta} \|^4
\big),
\end{equation}
\end{small}
with \eqref{indexResentation} for the representation of the index pair $(\alpha,\beta)$, and \eqref{indexResentationTilde} for $(\widetilde{\alpha},\widetilde{\beta})$.

Recalling \eqref{EdgeSetting}, and in \eqref{deterJ6} rewrite
\begin{small}
\begin{equation}\label{g3Est2}
(-1)^{\widetilde{\alpha}+\widetilde{\beta}}
\det\big(
U[I_1/\{ \widetilde{\alpha}\};I_1/\{ \widetilde{\beta} \}]
\big)=\det\big(
U[I_1] 
\big)\big(
U[I_1] 
\big)^{-1}_{\widetilde{\beta},\widetilde{\alpha}},
\end{equation}
\end{small}
and further
\begin{small}
\begin{equation}\label{g3Est3}
\sum_{\widetilde{\alpha},\widetilde{\beta}=1}^t
\big(
U[I_1]\overrightarrow{\eta_L}
\overrightarrow{\eta_L}^*U[I_1]
\big)_{\widetilde{\alpha},\widetilde{\beta}}
\big( U[I_1] \big)_{\widetilde{\beta},\widetilde{\alpha}}^{-1}
={\rm Tr}\big(
U[I_1]\overrightarrow{\eta_L}
\overrightarrow{\eta_L}^*
\big)=\overrightarrow{\eta_L}^*U[I_1]\overrightarrow{\eta_L}.
\end{equation}
\end{small}
Combining \eqref{FirstFactor}, \eqref{deterJ6}, \eqref{deterJ60}, \eqref{g3Est1}-\eqref{g3Est3}, we have
\begin{small}
\begin{multline}\label{g3 eta y Analysis}
g_3\big(
\overrightarrow{\eta},y
\big)=N|y|^{2t}\big(
\overrightarrow{\eta_L}^*U[I_1]\overrightarrow{\eta_L}
\big)\det\big(
U[I_1]
\big)\det\big(
U^{-1}[I_2]
\big)\big| \det(z_0\mathbb{I}_{r_0}-J_1) \big|^2
+O\big(
N\| \overrightarrow{\eta} \|\| \overrightarrow{\eta_R} \|
|y|^{2t}
\big)
\\
+\| \overrightarrow{\eta} \|^2
O\Big(\big(
|y|+N^{-\frac{1}{2}}+\| \overrightarrow{\eta} \|^2
\big)^{2t-2}+
N|y|\Big(
\big(
|y|+N^{-\frac{1}{2}}+\| \overrightarrow{\eta} \|^2
\big)^{2t-1}-|y|^{2t-1}
\Big)+N|y|^{2t+2}
\Big).
\end{multline}
\end{small}

Now combining \eqref{FirstFactor}, \eqref{g1 eta y Analysis}, \eqref{g2 eta y Analysis} and \eqref{g3 eta y Analysis}, we can finish the proof of Proposition \ref{getayexpan}.

\hspace*{\fill}

\hspace*{\fill}

 \noindent{\bf Acknowledgements}  
 We  would like to thank  Dang-Zheng Liu for useful comments and suggestions on this paper. This  work 
 was   supported by  the National Natural Science Foundation of China \# 12371157.

 \appendix
  
 \section{Proof of Equation  \eqref{PSJacobian}.}  \label{Appendix1}
By following almost the same procedure in \cite[Appendix B]{FK2007}, we can prove \eqref{PSJacobian}. While for completeness, we will still give a complete proof for \eqref{PSJacobian}.

 Take differential on both side of \eqref{PartialSchur}, we can get that
 \begin{small}
 \begin{equation}\label{Apendix11}
 {\rm d}X_N={\rm d}U\widetilde{X}_N U^*
 +U{\rm d}\widetilde{X}_NU^*+U\widetilde{X}_N{\rm d}U^*.
 \end{equation}
 \end{small}
 As $U^*U=\mathbb{I}_N$, we have $U^*{\rm d}U=-{\rm d}U^*U$, so we can denote the skew-Hermitian differential matrix
 \begin{small}
 \begin{equation}\label{dW differ}
 {\rm d}W=U^*{\rm d}U
 =\begin{bmatrix}
 {\rm d}a & -{\rm d}\overrightarrow{b}^*
 \\
{\rm d}\overrightarrow{b} & {\rm d}W_{N-1}
 \end{bmatrix},
 \end{equation}
 \end{small}
 here, from $U=[\overrightarrow{u_1},U_2(\overrightarrow{u_1})]$, we know
 \begin{small}
 \begin{equation}\label{d vec b}
 {\rm d}\overrightarrow{b}=U_2\big(
 \overrightarrow{u_1}
 \big)^*{\rm d}\overrightarrow{u_1}.
 \end{equation}
 \end{small}

Combining \eqref{PartialSchur}, \eqref{Apendix11} and \eqref{dW differ} we have 
\begin{small}
\begin{multline}\label{Apendix12}
{\rm d}\widehat{X}_N=U^*{\rm d}X_N U={\rm d}W \widetilde{X}_N-\widetilde{X}_N {\rm d}W
{\rm d}\widetilde{X}_N
\\
=\begin{bmatrix}
-\overrightarrow{\omega}^*{\rm d}\overrightarrow{b}
&
\overrightarrow{\omega}^*{\rm d}a-{\rm d}\overrightarrow{b}^*X_{N-1}
+z_n{\rm d}\overrightarrow{b}^*
-\overrightarrow{\omega}^*{\rm d}W_{N-1}
\\
(z_n\mathbb{I}_{N-1}-X_{N-1}){\rm d}\overrightarrow{b}
& {\rm d}\overrightarrow{b}\overrightarrow{\omega}^*
+{\rm d}W_{N-1}X_{N-1}-X_{N-1}{\rm d}W_{N-1}
\end{bmatrix}+
\begin{bmatrix}
{\rm d}z_n & {\rm d}\overrightarrow{\omega}^*
\\ & {\rm d}X_{N-1}
\end{bmatrix}.
\end{multline}
\end{small}
 
As $U$ is unitary, we have
\begin{small}
\begin{equation*}
\prod_{i,j=1}^N \Big(
{\rm d}\Re \big( \widehat{X}_{N} \big)_{i,j}
{\rm d}\Im \big( \widehat{X}_{N} \big)_{i,j}
\Big)=
\prod_{i,j=1}^N \Big(
{\rm d}\Re \big( X_{N} \big)_{i,j}
{\rm d}\Im \big( X_{N} \big)_{i,j}
\Big),
\end{equation*}
\end{small} 
 combining this and \eqref{Apendix12}, from basic computations we have
 \begin{small}
 \begin{equation}\label{Apendix13}
 [{\rm d}X_N]=\prod_{i,j=1}^N \Big(
{\rm d}\Re \big( X_{N} \big)_{i,j}
{\rm d}\Im \big( X_{N} \big)_{i,j}
\Big)=
\big|
\det(z_n\mathbb{I}_{N-1}-X_{N-1})
\big|^2[{\rm d}z_n]
[{\rm d}\overrightarrow{\omega}]
[{\rm d}\overrightarrow{b}]
[{\rm d}X_{N-1}].
 \end{equation}
 \end{small}
Note from \eqref{d vec b}, we know \begin{small}
$[{\rm d}\overrightarrow{b}]:=\prod_{i=1}^{N-1} \big(
{\rm d}\Re b_i
{\rm d}\Im b_i
\big)$
\end{small} is in fact the volume element ${\rm d}S\big( \overrightarrow{u_1} \big)$ on the surface of the unit sphere. Combining this and \eqref{Apendix13} we can finish the proof of \eqref{PSJacobian}.

  \section{1-point correlation function for the outlier eigenvalues}  \label{Appendix2}
In this section, for the random matrix model $X$ defined in \eqref{model}, we consider the following 1-point correlation function for the outlier eigenvalues:
\begin{equation}\label{1pointcorrelation}
R_N^{(1)}(A_0;z)=\sum_{i=1}^N\mathbb{E}\big(
\delta(z-z_i)
\big),
\end{equation}  
where $z_1,\cdots,z_N$ are eigenvalues of $X$. Now setting $\tau=1$, we consider the cases when $A_0$ takes the form \eqref{Jdetailoutlier1} and $ |z_0|^2>1 $. 
In this case, the perturbed matrix $X$ in definition \ref{GinU} will have outliers located around $z_0$ asymptotically, see e.g. \cite[Theorem 1.8]{BC16}. Here, by using \cite[Proposition 1.3]{LZ12} and the methodology in the proof of Theorem \ref{OverlapStaEdgeOutlier1}, we can derive the corresponding asymptotic behaviour for the 1-point correlation function at the outlier eigenvalues. Denote $\vec{e}_i$ the standard basis in $\mathbb{R}^r$, for $i=1,\cdots,r$.
\begin{proposition}\label{1pointStaEdgeOutlier1}
With the assumptions in \eqref{meanmatrix} and \eqref{Jdetailoutlier1}, let $r$ be finite and $\tau=1$. If $|z_0|^2>1$, 
 then 
as $N\to \infty$ the scaled 1-point correlation function 
\begin{multline}\label{1pointStaEdgeEquaOutlier1}
N^{-\frac{1}{r}}R_N^{(1)}\big(
A_0,z_0+N^{-\frac{1}{2r}}\hat{z}
\big)=\frac{r^2 |\hat{z}|^{2r-2}}{\pi}
\frac{1-|z_0|^{-2}}{\big(
\vec{e}_1^*P^*P \vec{e}_1
\big)
\big(
\vec{e}_r^*P^{-1}(P^*)^{-1} \vec{e}_r
\big)}
\\
\times
\exp\Big\{-
\big(
1-|z_0|^{-2}
\big)
\big(
\vec{e}_1^*P^*P \vec{e}_1
\big)^{-1}
\big(
\vec{e}_r^*P^{-1}(P^*)^{-1} \vec{e}_r
\big)^{-1}
\big|\hat{z}\big|^{2r}
\Big\}
+O\big(
N^{-\frac{1}{2r}}(1+|\hat{z}|)^{2r+1}
\big),
\end{multline}
uniformly for all $\hat{z}\in\mathbb{C}$ such that $|\hat{z}|\leq \log(N)$.
\end{proposition}
\begin{remark}
For the model \eqref{model}, the asymptotic positions of the outliers have been characterized exactly, and the relative microscopic eigenvalue statistics have been well understood, see \cite{Ta13,BC16,BR}.
\end{remark}
\begin{remark}
By using Proposition \ref{1pointStaEdgeOutlier1}, we can compute the expectation of number of eigenvalues for the perturbed matrix $X$ in certain compact region, which contains the outlier point $z_0$ and regresses to $z_0$ as $N\rightarrow+\infty$:
\begin{multline*}
\mathbb{E}\big(
\textit{Number of eigenvalues in }B(z_0,N^{-\frac{1}{2r}}\log(N))
\big)
\\
=\mathbb{E}\Big(
\sum_{i=1}^N
\int_{B(z_0,N^{-\frac{1}{2r}}\log(N))}
\delta(z-z_i){\rm d}z
\Big)
\\
=
\int_{B(z_0,N^{-\frac{1}{2r}}\log(N))}
R_N^{(1)}\big(
A_0,z\big){\rm d}z
\\
=
\int_{B(0,\log(N))}
N^{-\frac{1}{r}}R_N^{(1)}\big(
A_0,z_0+N^{-\frac{1}{2r}}\hat{z}\big){\rm d}\hat{z}
\\
=
\int_{B(0,\log(N))}
\Big(
\frac{r^2 |\hat{z}|^{2r-2}}{\pi}
\frac{1-|z_0|^{-2}}{\big(
\vec{e}_1^*P^*P \vec{e}_1
\big)
\big(
\vec{e}_r^*P^{-1}(P^*)^{-1} \vec{e}_r
\big)}
\\
\times
\exp\Big\{-
\big(
1-|z_0|^{-2}
\big)
\big(
\vec{e}_1^*P^*P \vec{e}_1
\big)^{-1}
\big(
\vec{e}_r^*P^{-1}(P^*)^{-1} \vec{e}_r
\big)^{-1}
\big|\hat{z}\big|^{2r}
\Big\}
+O\big(
N^{-\frac{1}{2r}}(1+|\hat{z}|)^{2r+1}
\big)
\Big){\rm d}\hat{z}
\\
=r+O\big(
N^{-\frac{1}{2r}}(log(N))^{2r+3}
\big).
\end{multline*}
\end{remark}
 Therefore, under the assumptions in \eqref{meanmatrix} and \eqref{Jdetailoutlier1} and with $|z_0|^2>1$, the expectation of number of eigenvalues of $X$ in $B(z_0,N^{-\frac{1}{2r}}\log(N))$ is $r$ asymptotically.

\end{document}